\documentclass [12pt]{article}
\usepackage{amsmath,amssymb,cite}

\usepackage[english]{babel}
\usepackage[dvips]{graphicx}

\usepackage{indentfirst}
\numberwithin{equation}{section}

\begin{document}
\begin{flushright}
DIAS-04-06\\
\end{flushright}
\begin{center}
{\large{\bf A gauge invariant UV-IR mixing and the corresponding phase transition for $U(1)$ fields on the Fuzzy Sphere}}\\
\bigskip
P.Castro-Villarreal$^{*+}$ , R.Delgadillo-Blando$^{*+}$ , Badis Ydri$^{*}$.
\bigskip

$^{*}${\it School of Theoretical Physics, \\
Dublin Institute for Advanced Studies, Dublin, Ireland.}\\
$^{+}${\it Dept. de Fisica, Centro de  Investigaciones  de Estudios Avanzados\\
del IPN, Apdo. Postal 14-740, 07000, Mexico D.F., Mexico.}\\

\end{center}

\begin{abstract}
From a string theory point of view the most natural gauge action on the fuzzy sphere ${\bf S}^2_L$ is the  Alekseev-Recknagel-Schomerus action which is a particular combination of the Yang-Mills action and the Chern-Simons term . The differential calculus on the fuzzy sphere is $3-$dimensional and thus the field content of this model consists of a $2$-dimensional gauge field together with a scalar fluctuation normal to the sphere . For $U(1)$ gauge theory we compute the quadratic effective action and shows explicitly that the tadpole diagrams and the vacuum polarization tensor contain a gauge-invariant UV-IR mixing in the continuum limit $L{\longrightarrow}{\infty}$ where $L$ is the matrix size of the fuzzy sphere. In other words the quantum $U(1)$ effective action does not vanish in the commutative limit and a noncommutative anomaly survives . We compute the scalar effective potential and prove the gauge-fixing-independence of the limiting model $L={\infty}$ and then show explicitly that the one-loop result predicts  a first order phase transition which was observed recently in simulation . The one-loop result for the $U(1)$ theory is exact in this limit . It is also argued that if we add a large mass term for the scalar mode the UV-IR mixing will be completely removed from the gauge  sector . It is  found in this case to be confined to the scalar sector only. This is in accordance with the large $L$ analysis of the model . Finally we show that the phase transition becomes harder to reach starting from small couplings when we  increase $M$ . 
\end{abstract}

\section{Introduction and results}
The fuzzy sphere ${\bf S}^2_L$ as an approximation of the ordinary
sphere is due originally  to Madore \cite{Madore}. See also Hoppe \cite{hoppe}. Unlike naive
lattice prescription this approximation preserves all symmetries
of the continuum theory such as rotational symmetry of the ordinary
sphere, local gauge symmetry of the standard model \cite{badis1,badis,badis2}
and most notably  supersymmetry \cite{sechkin}. In particular it
is shown that chiral symmetry is maintained without fermion
doubling \cite{badis2,fmd} and that this approximation captures  most
of the topology of the original commutative sphere such as
monopole configurations, integer winding numbers and the index
theorem \cite{mono,newtop}.

It is believed that field theories on continuum manifolds  can
always be regularized in this fashion, i.e by replacing the
underlying space with a finite dimensional (fuzzy) matrix model.
Extension to $4$ dimensions for example entails the use of either
$a)$ the Cartesian product of two fuzzy spheres ${\bf
S}^2{\times}{\bf S}^2$ or $b)$ a fuzzy ${\bf CP}^2$ 
\cite{ydri,cp}. Fuzzy ${\bf S}^4$ as obtained from squashed ${\bf CP}^3$ is also 
a candidate  for $4-$dimensional fuzzy physics  \cite{juliete}. From a practical point of view the spaces ${\bf S}^2_L$, ${\bf S}^2_L{\times}{\bf R}$, ${\bf S}^2_L{\times}{\bf R}^2$ and ${\bf S}^2_L{\times}{\bf S}^2_L$ are the most useful  for analytical manipulation since clearly they will only involve the well known $SU(2)$ Clebsch-Gordan coefficients \cite{ydri}. We note in passing that other higher dimensional fuzzy spaces can also be formulated \cite{others}.

The motivation for studying quantum field theories on fuzzy models
is therefore two-fold. Firstly this is clearly a novel way of (possibly)
simulating ordinary gauge theories, QCD in particular, based on
random matrix models which is potentially superior to current
methods because of the symmetry-topology arguments  outlined
above. Secondly fuzzy spaces because of their close connection to
Moyal-Weyl noncommutative spaces  could provide a systematic way
of regularizing and then renormalizing Moyal-Weyl quantum field
theories. Indeed and as it turns out field theories on the noncommutative
Moyal-Weyl spaces can also be regularized by replacing them with
finite dimensional (fuzzy) matrix models \cite{badis,ydri}. In either cases the limit
of interest is a continuum large $L$ limit where $L$ is the size
of the matrices approximating (say) in two dimensions ${\bf S}^2$
or ${\bf R}^2_{\theta}$. 

It is well known that perturbation theory of fuzzy scalar models
are plagued by the so-called UV-IR mixing. On fuzzy spaces the mixing is defined by the requirement  that the fuzzy
quantum actions do not approach the corresponding quantum effective actions on the commutative spaces \cite{UVIR,ydri}. We remark that this criterion for the existence 
of the UV-IR mixing on fuzzy spaces is different from the criterion on the noncommutative Moyal-Weyl planes found in \cite{MinwallaSeiberg}. However the
fuzzy UV-IR mixing  can be viewed as a regularized version of the
UV-IR mixing on the noncommutative plane which will reduce to it in some appropriate flattening limit. This was shown explicitly for ${\lambda}{\phi}^4$ theory in $2$ and $4$ dimensions in  
\cite{uvir,ydri}.  

The presence of this mixing on the
fuzzy sphere is however a major problem from a theoretical point
of view since it means that the scalar model on the fuzzy sphere
does not really  approximate (as it should) the corresponding
scalar model on the ordinary sphere. A priori any simulation of
such fuzzy models will therefore give wrong results. Several ways
of dealing with this problem were devised \cite{UVIR,ydri} and a numerical study to
probe the nonperturbative properties of the model was undertaken
in \cite{martin}.

In this article we will study $U(1)$ gauge theory on the fuzzy sphere $S^{2}_{L}$. The main results are summarized as follows. The differential calculus on the fuzzy sphere is three dimensional and as a consequence a spin 1 vector field $\vec{A}$ is intrinsically $3-$dimensional. Each component $A_a$, $a=1,2,3$, is an element of some matrix algebra $Mat_{L+1}$. Thus $U(1)$ symmetry will be implemented by $U(L+1)$ transformations $U$ as follows $A_a{\longrightarrow}UA_aU^{+}+U[L_a,U^{+}]$ where $L_a$ are the  generators of $SU(2)$ in the irreducible representation $\frac{L}{2}$ of the group. 

On the fuzzy sphere $S^2_L$ it is not possible to split the vector field $\vec{A}$ in a gauge-covariant fashion into a tangent two-dimensional gauge field and a normal scalar fluctuation. This splitting is done trivially on the commutative sphere by simply writing $A_a=n_a{\Phi}+a_a$ where $\vec{n}$ is the unit vector on $S^2$, $\Phi=\vec{n}\cdot \vec{A}$ is the normal gauge-invariant component of $\vec{A}$ and $\vec{a}$ is the tangent gauge field. However although we do not have the analogue of $a_a$ on the fuzzy sphere we can still write a gauge-covariant expression for the normal scalar component in terms of $A_a$ which reads 
\begin{eqnarray}
\Phi=\frac{1}{2}\left(x_aA_a+A_ax_a+\frac{A_a^2}{\sqrt{L_a^2}}\right),
\end{eqnarray}
where $x_a=\frac{L_a}{\sqrt{L_a^2}}$ are the matrix coordinates on fuzzy $S^2_L$. In the limit $L\to\infty$ it is not difficult to see that the matrix coordinates $x_a$ tend to the commutative coordinates $n_a$ and the scalar field $\Phi$ tends to $\vec{n}\cdot \vec{A}$. 

The most general action (up to quartic power in $A_a$) which is invariant under $U(1)$ transformations  on the fuzzy sphere $S^2_L$ is given by 

\begin{eqnarray}
S_L[A_a]&=&-\frac{1}{4g^2}Tr_L\left[F_{ab}^{(0)}+[A_a,A_b]\right]^2-\frac{i}{2g^2}{\epsilon}_{abc}Tr_L\left[\frac{1}{2}F_{ab}^{(0)}A_c+\frac{1}{3}[A_a,A_b]A_c\right]
\nonumber\\
&+&\frac{2M^2}{g^2}Tr_L \Phi^2
+\frac{{\alpha}|L|}{g^2}Tr_L \Phi.\label{main0}
\end{eqnarray}
$F_{ab}=F_{ab}^{(0)}+[A_a,A_b]$ is the $U(1)$ covariant curvature where $F_{ab}^{(0)}=[L_a,A_b]-[L_b,A_a]-i{\epsilon}_{abc}A_c$. In the continuum limit $L{\longrightarrow}{\infty}$ all commutators vanish and we get the action 
\begin{eqnarray}
S_{\infty}[A_a]&=&-\frac{1}{4g^2} \int_{S^2} \frac{d{\Omega}}{4{\pi}}(F_{ab}^{(0)})^2-\frac{i}{2g^2}{\epsilon}_{abc}\int_{S^2}\frac{d{\Omega}}{4\pi}\frac{1}{2}F_{ab}^{(0)}A_c+\frac{2M^2}{g^2}\int_{S^2}\frac{d{\Omega}}{4\pi} \Phi^2\nonumber\\
&+&\frac{{\alpha}|L|}{g^2}\int_{S^2} \frac{d{\Omega}}{4{\pi}}\Phi.\label{main}
\end{eqnarray}
$F_{ab}^{(0)}$ becomes the $U(1)$ curvature which is now given by $F_{ab}^{(0)}={\cal L}_aA_b-{\cal L}_bA_a-i{\epsilon}_{abc}A_c$ where ${\cal L}_a=-i{\epsilon}_{abc}n_b\frac{\partial}{{\partial}n_c}$. For $U(1)$ theory this curvature is exactly gauge invariant. 

The continuum action $S_{\infty}$ is at most quadratic in the field $A_a$ (which can therefore be integrated out easily in the path integral) and as a consequence the corresponding effective action will be essentially given by $S_{\infty}$ itself. On the other hand the quantization of the fuzzy action $S_L$ is much more involved and yields a non-trivial effective action. As it turns out the continuum limit of this fuzzy effective action does not tend to $S_{\infty}$ for generic values of the parameters $M$ and $\alpha$. This is the signature of the UV-IR mixing in this model. In this article we computed explicitly {\it the quadratic } effective action for the values $M=\alpha=0$ and found it to be given in the continuum limit $L{\longrightarrow}{\infty}$ by the expression

\begin{eqnarray}
{\Gamma}_{2}&=& -\frac{1}{4g^2}\int
\frac{d{\Omega}}{4{\pi}}F_{ab}^{(0)}(1+2g^2{\Delta}_3)F_{ab}^{(0)}-\frac{i}{4g^2}{\epsilon}_{abc}\int
\frac{d{\Omega}}{4{\pi}}F_{ab}^{(0)}(1+2g^2{\Delta}_3)A_c+4|L|\int\frac{d{\Omega}}{4{\pi}}\Phi \nonumber\\
&+&{\rm non~local~ quadratic ~terms}.\label{main1}
\end{eqnarray}
The operator ${\Delta}_3$ is a complicated function of the Laplacian ${\cal L}^2$ which is defined in equation (\ref{introduction}). The $1$  in $1+2g^2{\Delta}_3$ corresponds to the classical action whereas ${\Delta}_3$ is the quantum correction. By comparing (\ref{main}) and (\ref{main1}) it is clear that $1+2g^2{\Delta}_3$ provides a non-local renormalization of the inverse coupling constant $1/g^2$ whereas the third term in (\ref{main1}) provides a local renormalization of the coupling constant $\alpha$ which acquires the value $4g^2$. The last terms in (\ref{main1}) are new non-local quadratic terms which have no counterpart in the classical action. Their explicit expression is given in (\ref{46}). Remark that the quadratic action (\ref{main1}) is already gauge-invariant which will not be the case for $U(n)$ theories. We have thus established the existence of a gauge-invariant UV-IR mixing problem in $U(1)$ gauge theory on fuzzy $S^2_L$ for the values $M=\alpha=0$. It is only natural to expect that the same result will also hold for generic values of the parameters $M$ and $\alpha$.

In this paper we will also show that this UV-IR mixing problem is only confined to the scalar sector of the model in the following sense. We consider the model (\ref{main0}) for $\alpha=0$ and finite $M$. We remark that at the level of the classical continuum action (\ref{main}) the limit $M{\longrightarrow}{\infty}$ projects out the scalar fluctuation $\Phi$. Indeed in this limit this field becomes infinitely heavy and thus decouples from the rest of the dynamics.  If we decide to quantize the model (\ref{main0}) and then take the limit $M{\longrightarrow}\infty$ first and then $L{\longrightarrow}{\infty}$  then one finds that the quantum corrections depend only on the scalar field $\Phi$ [see equation (\ref{rodrigo})]. Hence in this limit the effective action of the two-dimensional gauge field seems to be given essentially by the classical action whereas the normal scalar field still gets non-trivial quantum contributions in the path integral due to the underlying noncommutativity. This is consistent with the case of pure scalar models studied in \cite{uvir,UVIR,ydri} but the detail structure of the remaining UV-IR mixing in here is different.

A more elegant test for the UV-IR mixing in this theory can be given in terms of the normal scalar field $\Phi$. Let us consider the following simple normal field configuration defined by
\begin{eqnarray}
A_a=\left(\phi-1\right)L_a.\label{phi}
\end{eqnarray} 
The normalization is chosen for latter convenience. The real number $\phi$ is related to the normal scalar field $\Phi$ by $\phi=\sqrt{1+2\Phi/\sqrt{L_a^2}}$. 

It is a trivial exercise to compute the classical action (\ref{main0}) for this configuration and one obtains the classical potential given by 
 
\begin{eqnarray}
S=\frac{\sqrt{L_a^2}}{2g^2}\left[\phi^4-\frac{4}{3}\phi^3+M^2\left(\phi^2-1\right)^2+\alpha {\phi}^2\right].
\end{eqnarray}
The {\it full} effective action in the continuum large $L$ limit is given by
\begin{eqnarray}
{\Gamma}=S + 4\sqrt{L_a^2}\log\phi.
\end{eqnarray}
For $M=0$ and $\alpha =0$ the classical potential has a minimum at ${\phi}=1$ for which the above normal gauge field $A_a$ vanishes. This is the vacuum of the classical theory. To find the effect of the quantum corrections on this vacuum we take the first and second derivatives of the effective potential ${\Gamma}$ with respect to $\phi$. The condition ${\Gamma}^{'}=0$ will give us extrema of the model whereas the condition ${\Gamma}^{''}=0$ tells us when we go from bounded potential (a minimum) to unbounded potential. Solving the above two equations yield immediately the minimum ${\phi}_{*}=\frac{3}{4}$ with the corresponding critical value
\begin{eqnarray}
g_{*}^2=\frac{1}{8}\left(\frac{3}{4}\right)^3.\label{critical}
\end{eqnarray}
We can  conclude from this result that at the critical value (\ref{critical}) a first order phase transition occurs which separates the fuzzy sphere phase where $\phi$ has a well defined minimum from the pure matrix phase where the minimum disappears. In the fuzzy sphere phase the interpretation of a $U(1)$ gauge theory on a sphere is valid and it holds for $g<g_{*}$. The matrix phase is where this interpretation brakes down and it holds for $g>g_{*}$. 
This agrees nicely with the result of \cite{jun} which was however obtained by simulating the full theory. In other words the one-loop result obtained here is exact.

For $\alpha=0$ and in the limit $M{\to}{\infty}$ we can also compute the values of $\phi$ and $g^2$ at the critical point and find them to be given by ${\phi}_{*}{\sim}{\pm}1/{\sqrt{2}}$ and $g_{*}^2{\sim}M^2/8$. In other words the phase transition happens each time at a larger value of the coupling constant when $M$ is increased and hence it is harder for the system to reach the pure matrix phase for large enough masses if one starts from the fuzzy sphere phase.

This article  is organized as follows. Section $2$ introduces the fuzzy sphere. In section $3$ the fuzzy gauge field is defined and its different actions are written down. In section $4$ we quantize the model in the Feynman-'t Hooft Background field gauge and compute the quadratic effective action by computing tadpole graphs and the vacuum polarization tensor. In section $5$ we study the continuum limit of the theory in great detail and show the existence of a gauge-invariant UV-IR mixing in the limit. In section $6$ we compute the effective potential of the scalar mode and show the presence of a first order phase transition in the model. The critical point in the strict limit $L={\infty}$ is written down and we show its gauge-fixing independence. In section $7$ we study the large mass limit of the model and show that in this limit the scalar mode is decoupled from the gauge modes and correspondingly the gauge-sector of the theory is UV-IR free in the continuum limit. We also show that in the presence of a large mass term the first order phase transition of the model  becomes harder to reach from small couplings as we increase the mass. In the appendices we give the detail of our calculation.

\section{The Fuzzy Sphere}

In here we define the non-commutative fuzzy sphere by Connes spectral triple $(Mat_{L+1},{H}_L,{\Delta}_L)$ \cite{connes}. $Mat_{L+1}$ is the algebra of $(L+1){\times}(L+1)$ matrices which acts on an $(L+1)-$dimensional Hilbert space ${H}_L$ with inner product $(M,N)=\frac{1}{L+1}Tr(M^{\dagger}N)$ where $M,N{\in}Mat_{L+1}$. ${\Delta}_L$ is the Laplacian on the fuzzy sphere which we will define shortly. Matrix coordinates on ${\bf S}^2_L$ are defined by

\begin{eqnarray}
x_1^2+x_2^2+x_3^2=1~,~
[x_a,x_a]=\frac{i}{|L|}{\epsilon}_{abc}x_c,
\end{eqnarray}
with
\begin{equation}
x_a=\frac{L_a}{|L|}.\label{22}
\end{equation}
$L_a$ are the generators of the irreducible representation $\frac{L}{2}$ of $SU(2)$, i.e  $[L_a,L_b]=i{\epsilon}_{abc}L_c$, $\sum_a L_a^2=|L|^2{\equiv}\frac{L}{2}(\frac{L}{2}+1)$.  The Hilbert space $H_L$ is naturally associated with this representation. These definitions are motivated by the fact that ${\bf S}^2$ is nothing but the co-adjoint orbit $SU(2)/U(1)$ which is thus a symplectic manifold and hence it can be quantized in a canonical fashion by simply quantizing the volume form ${\omega}=dcos{\theta}{\wedge} d{\phi}$ \cite{badis}.

``Fuzzy'' functions on ${\bf S}^2_L$ are linear operators in the
matrix algebra while derivations are inner defined by the
generators of the adjoint action of $SU(2)$, in other words the
derivative of the fuzzy function ${\phi}{\in}Mat_{L+1}$ in the
space-time direction $a$ is the commutator $[L_a,{\phi}]$. This
can also be put in the form
\begin{eqnarray}
AdL_a({\phi})\equiv[L_a,{\phi}]=(L_a^L-L_a^R)({\phi}){\equiv}{\cal L}_a({\phi}),\label{23}
\end{eqnarray}
where $L_a^L$'s and $-L_a^R$'s are the generators of the IRR $\frac{L}{2}$ of
$SU(2)$ which act respectively on the left and on the right of the algebra
$Mat_{L+1}$, i.e $L_a^L{\phi}{\equiv}L_a{\phi}$,
$L_a^R{\phi}{\equiv}{\phi}L_a$ for any
${\phi}{\in}Mat_{L+1}$.

A natural choice of the Laplacian operator ${\Delta}_L$ on the fuzzy sphere is therefore given by the following Casimir operator
\begin{eqnarray}
{\Delta}_L=(L_a^L-L_a^R)^2{\equiv}{\cal L}^2.\label{24}
\end{eqnarray}
Thus the algebra of matrices $Mat_{L+1}$ decomposes under the
action of the group $SU(2)$ as
$\frac{L}{2}{\otimes}\frac{L}{2}=0{\oplus}1{\oplus}2{\oplus}..{\oplus}L$
(The first $\frac{L}{2}$ stands for the left action of the group
while the other $\frac{L}{2}$ stands for the right action). It
is not difficult to convince ourselves that this Laplacian has a
cut-off spectrum of the form $k(k+1)$ where $k=0,1,...,L$. As a
consequence a general function on ${\bf S}^2_L$, i.e scalar
fields, can be expanded in terms of polarization tensors as
follows
\begin{eqnarray}
{\phi}=\sum_{k=0}^{L}\sum_{m=-k}^{k}{\phi}_{km}\hat{Y}_{km}.
\end{eqnarray}
For an extensive list of the properties of
$\hat{Y}_{km}(l)$ see \cite{VKM}.

\section{Classical gauge fields on ${\bf S}^2_L$}
\subsection{The Alekseev-Recknagel-Schomerus action. } 

It was shown in \cite{ars} that the dynamics  of open strings
moving in a curved space with ${\bf S}^3$ metric in the presence
of a non-vanishing Neveu-Schwarz B-field   and with Dp-branes is
equivalent to leading order in the string tension to a gauge
theory on a noncommutative fuzzy sphere with a Chern-Simons term. 
The full $U(n)$ action on the fuzzy sphere they found is given
by the combination

\begin{eqnarray}
S_{ARS}[D_a]&=&S_{YM}[D_a]+S_{CS}[D_a],
\end{eqnarray}
where $D_a$ is the covariant derivative with curvature
$F_{ab}=[D_a,D_b]-i{\epsilon}_{abc}D_c$ and the Yang-Mills and
Chern-Simons actions are given respectively by
\begin{eqnarray}
S_{YM}[D_a]&=&-\frac{1}{4g^2} Tr_{L}tr~~F_{ab}^2\nonumber\\
S_{CS}[D_a]&=&-\frac{1}{6g^2}Tr_{L}tr\left[i\epsilon_{abc}F_{ab}D_c+(D_a^2-L_a^2)\right].
\end{eqnarray}
This result is simply an extension of the original result of
\cite{sw} in which strings moving in a flat space in the presence
of a constant N-S B-field  are described in the limit
${\alpha}^{'}{\longrightarrow}0$ by a Moyal-Weyl noncommutative
gauge theory. From string theory  point of view the above
combination of Yang-Mills and Chern-Simons actions is therefore
the most natural candidate for a gauge action on the fuzzy sphere. 
In most of this paper we will thus work with the action

\begin{eqnarray}
S_{ARS}[D_a]=\frac{1}{g^2}Tr_{L}tr\left[
-\frac{1}{4}[D_a,D_b]^2+\frac{i}{3}\epsilon_{abc}[D_a,D_b]D_c\right]
+\frac{1}{6g^2}Tr_{L}tr|L|^2.\label{action}
\end{eqnarray}
Gauge transformations are implemented here by the unitary
transformations $U\in U_{n\left(L+1\right)}$ as follows
$D_a{\rightarrow}D_a^{\prime}=UC_{a}U^{-1}$. In above we have adopted the
convention $Tr_{L}=\frac{Tr}{L+1}$ and set $R=1$ where $R$ is the
radius of the sphere . $tr$ is the trace over the gauge group.
$S_{ARS}$ is the Alekseev-Recknagel-Schomerus action.

The first remark about this action is the fact that there is no
quadratic term, i.e the term $D_a^2$ from the YM part cancels exactly
the term $D_a^2$ from the CS. Furthermore we remark (as we will
show shortly) that in the Feynman gauge $\xi=1$, the kinetic
term reduces to ${\cal L}^2$: This is simply the inverse
propagator in the plane which can already be seen at the level of
equations of motion. Indeed varying the action yields

\begin{equation}
{\delta} S_{ARS} = -\frac{1}{g^2}Tr_Ltr{\delta}D_a
\left([D_b,[D_a,D_b]]-i\epsilon_{abc}[D_b,D_c]\right),
\end{equation}
and thus we obtain the equations of motion
\begin{equation}
[D_b,F_{ab}]=0~,~F_{ab}=[D_a,D_b]-i{\epsilon}_{abc}D_c.\label{EOM}
\end{equation}
As it was shown in \cite{ars} classical solutions in the presence
of the Chern-Simons term which are also absolute minima of the
action are characterized by $SU(2)$ IRR. (\ref{EOM}) can also be solved with general $SU(2)$ representations as well as with diagonal matrices.

A final remark about the action (\ref{action}) is to note that it
has the extra symmetry $D_a{\longrightarrow}D_a+{\alpha}_a{\bf
1}_{n(L+1)}$ where ${\alpha}_a$ are constant real numbers . This
symmetry needs to be fixed by restricting the covariant
derivative $D_a$ to be traceless, i.e by removing the zero mode
\cite{japan1,japan}. This symmetry manifests itself also in the
form $A_a{\longrightarrow}A_a+{\alpha}_a{\bf 1}_{n(L+1)}$ where
$A_a$ is the gauge field defined by $D_a=L_a+A_a$. Remark
however that for $D_a=B_a{\bf 1}_{n(L+1)}$ the action takes the
value $\frac{1}{6g^2}n|L|^2$ whereas for $A_a=B_a{\bf 1}_{n(L+1)}$
the action is identically zero.

As it turns out the action (\ref{action}) on its own does not
describe in the continuum limit pure gauge fields. Indeed we can
show that in the continuum limit the gauge field $A_a$ decomposes
as $A_a=a_a+n_a\phi$ where $a_a$ is the field tangent to the
sphere while $\phi n_{a}$ is the normal component  and as a
consequence the gauge action becomes
\begin{eqnarray}
S_{ARS}=-\frac{1}{4g^2}\int_{{\bf S}^2}
\frac{d{\Omega}}{4{\pi}}tr\bigg[f_{ab}^2+4i{\epsilon}_{abc}f_{ab}n_c\phi
+2[{\cal L}_a+a_a,\phi]^2-4{\phi}^2\bigg].\label{higgs}
\end{eqnarray}
$f_{ab}$ is clearly the curvature of the field $a_a$, i.e
$f_{ab}={\cal L}_aa_b-{\cal L}_ba_a-i{\epsilon}_{abc}a_c+[a_a,a_b]$. 
As one can immediately
see this theory consists of a $2-$dimensional gauge field with a
Higgs particle.
\subsection{Scalar action.} 

Next we show how to
suppress the scalar fluctuation $\phi$ in order to reduce the
model to a purely two-dimensional Yang-Mills theory. It is
obvious that the $3-$component gauge field is an element of the
full projective module ${Mat}_{n(L+1)}{\otimes}C^3$. This is due
to the fact that our description of the limiting commutative
sphere uses global coordinates $n_a$, $a=1,2,3$  instead of
local patches. Nevertheless strictly two-dimensional gauge
fields can still  be defined in the fuzzy as elements of the
tangent projective module $P^{T}(Mat_{n(L+1)}{\otimes}C^3)$ where
$P^{T}$ is the projector given by
\begin{eqnarray}
P^{T}_{ab}&=&{\delta}_{ab}-x_a x_b.\label{projector}
\end{eqnarray}
The meaning of this projector can be explained as follows. The
algebra of matrices $Mat_{n(L+1)}$ represents both the space, 
i.e the sphere, and the gauge group $U(n)$. It is clear
therefore that in the presence of a spin $1$ field the algebra of
matrices $Mat_{L+1}$ decomposes under the action of the rotation
group $SU(2)$ as follows
\begin{eqnarray}
{\Gamma}_{\frac{L}{2}}{\otimes}{\Gamma}_{\frac{L}{2}}{\otimes}{\Gamma}_1={\Gamma}_{\frac{L}{2}}{\otimes}\bigg({\Gamma}_{\frac{L}{2}+1}{\oplus}{\Gamma}_{\frac{L}{2}}{\oplus}{\Gamma}_{\frac{L}{2}-1}\bigg).
\end{eqnarray}
The two IRR ${\Gamma}_{\frac{L}{2}}$ stand for the left and right
actions of the group on the algebra $Mat_{L+1}$ whereas
${\Gamma}_1$ stands for the spin $1$ structure we want to add. 
It is rather a trivial exercise to compute the projectors on the
spaces ${\Gamma}_{\frac{L}{2}+1}$, ${\Gamma}_{\frac{L}{2}-1}$ and
verify that $P^T=P_{+}+P_{-}$. In other words $P_{+}A$ and
$P_{-}A$ are the components of the gauge field tangent to the
sphere whereas the normal component can be simply defined by
$P_{0}A$ where $P_{0}{\equiv}P^N=1-P^T$ which clearly projects on
the space ${\Gamma}_{\frac{L}{2}}$ and reads in terms of
components
\begin{eqnarray}
P^N_{ab}=x_ax_b.\label{normal}
\end{eqnarray}
By analogy with the language of continuum manifolds one can think
of $P^T$ as the projector onto the fuzzy tangent bundle (In fact
in the continuum limit $P^T$ is precisely the projector onto the
tangent bundle ${\bf T}{\bf S}^2$). This projector clearly
satisfies $(P^T)^2=P^T=P^{T+}$, $P^{T}_{ab}x_b=0$,
$x_aP^{T}_{ab}=0$ and  $P^{T}_{ab}P^{T}_{ba}=2$ which translates
the fact that $P^T_{ab}A_b$ is indeed a $2-$dimensional gauge
field.

A more practical way of implementing the projection $P^T$ is to constrain in a gauge-covariant way the gauge field $A_{a}$ to satisfy an extra condition and as a consequence reduce the number of its independent components from $3$ to $2$ . We adopt here the prescription of \cite{Nair}, i.e we impose on the gauge
field $A_{a}$ the gauge-covariant condition
\begin{eqnarray}
D_aD_a=|L|^2.\label{local}
\end{eqnarray}
This means that on the fuzzy sphere we allow only gauge
configurations $D_a$ which themselves live on a sphere of radius
$|L|$ in order for the model to describe a two-dimensional
Yang-Mills theory. This constraint reads explicitly
\begin{eqnarray}
{\Phi}=\frac{1}{2}\left(x_aA_a+A_ax_a+\frac{A_a^2}{|L|}\right)=0,\label{scalar}
\end{eqnarray}
and thus it is not difficult to check that in the continuum limit $L{\longrightarrow}{\infty}$ the normal component of the gauge field is zero, i.e ${\Phi}{\equiv}\vec{n}.\vec{A}=0$. In fact (\ref{scalar}) is the correct definition of the normal scalar field on the fuzzy sphere which is only motivated by gauge invariance. Following \cite{stein1} we incorporate the constraint (\ref{scalar}) into the theory by adding the following scalar action
\begin{eqnarray}
S_{M}[D_a]=\frac{1}{2g^2}\frac{M^2}{|L|^2}Tr_{L}tr(D_a^2-|L|^2)^2=\frac{2M^2}{g^2}Tr_{L}tr{\Phi}^2,\label{s1}
\end{eqnarray}
to (\ref{action}) where $M$ is a large mass. This term in the
continuum theory changes the mass term of the Higgs particle
appearing in (\ref{higgs}) from $\sqrt{2}$ to $\sqrt{2(1+2M^2)}$
and hence in the large $M$ limit the normal scalar field simply
decouples. The limit of interest in the remainder of this paper
is therefore $M{\longrightarrow}{\infty}$ first  then
$L{\longrightarrow}{\infty}$.

The most general $U(n)$ gauge action on the fuzzy sphere which is
at most quartic in the fields is therefore given by
\begin{eqnarray}
S[D_a]&=& S_{ARS}[D_a]+S_{M}[D_a]+\frac{\alpha}{2g^2}
Tr_Ltr(D_a^2-|L|^2)-\frac{1}{6g^2}Tr_Ltr|L|^2
\end{eqnarray}
or equivalently
\begin{eqnarray}
S&=&\frac{1}{g^2}Tr_{L}tr\left[-\frac{1}{4}[D_a,D_b]^2+\frac{i}{3}\epsilon_{abc}[D_a,D_b]D_c\right]
+\frac{1}{2g^2}\frac{M^2}{|L|^2}Tr_Ltr(D_a^2-|L|^2)^2\nonumber\\
&+&\frac{\alpha}{2g^2}Tr_Ltr(D_a^2-|L|^2),\label{actionbeta}
\end{eqnarray}
where we have also added a linear term in the scalar field with
parameter $\alpha$ while the constant is added so that the action
vanishes for pure gauges $D_a=UL_aU^{+}$.

\section{The Feynman-'t Hooft background field gauge}
\subsection{The effective action.} 
 
 The partition function of the theory depends therefore on $4$
parameters, the Yang-Mills coupling constant $g$, the mass $M$
of the normal scalar field, the linear coupling constant $\alpha$
as well as the size of the fuzzy sphere $L$, viz
\begin{eqnarray}
Z_{L}\left[J\right]\equiv Z_{L}\left[J;g,M,\alpha\right]=\int {\prod}_{a=1}^3\left[dC_a\right]e^{-S[C]-\frac{1}{g^2}Tr_LJ_aC_a}.
\end{eqnarray}
The equations of motion derived from the action (\ref{actionbeta}) read in terms of the gauge-covariant current $J_a$ as follows
\begin{eqnarray}
[C_b,F_{ab}]=\frac{M^2}{|L|^2}[C_a,C_b^2-|L|^2]_{+}+\alpha
C_a+J_a~.\label{eom}
\end{eqnarray}
Remark that the above action (\ref{actionbeta}) for $M{\neq}0$
and/or $\alpha{\neq}0$ does not enjoy the symmetry
$C_a{\longrightarrow}C_a+{\alpha}_a{\bf 1}_{n(L+1)}$ and thus we
can not simply remove the zero mode in this model.

 Now we adopt the background field method to
the problem of quantization of the $U(1)$ theory and then extend
the results to the $U(n)$ theory in a later communication. This
method consists in making a perturbation of the field around the
classical solution and then quantizing the fluctuation. Towards
this end we first separate the field as  $ C_a=D_a+Q_a$ and write
the action in the form
\begin{eqnarray}
S[C_a]&=&S[D_a]+\frac{1}{g^2}Tr_L\big(\hat{J}_a-J_a\big)Q_a
-\frac{1}{2g^2}Tr_L[D_a,Q_b]^2+ \frac{1}{2g^2}(1+\frac{M^2}{|L|^2})Tr_L[D_a,Q_a]^2\nonumber\\
&+&\frac{1}{g^2}Tr_LQ_a[F_{ab},Q_b]+\frac{1}{2g^2}Tr_LQ_a\bigg[\alpha{\delta}_{ab}+\frac{2M^2}{|L|^2}(D_c^2-|L|^2){\delta}_{ab}+\frac{4M^2}{|L|^2}D_aD_b\bigg]Q_b\nonumber\\
&+&({\rm
higher~ order~ terms~ in~ Q_a}) \label{action1}
\end{eqnarray}
where $\hat{J}_a=-
[D_b,F_{ab}]+\frac{M^2}{|L|^2}[D_a,D_b^2-|L|^2]_{+}+\alpha
D_a+J_a$, $F_{ab}=[D_a,D_b]-i{\epsilon}_{abc}D_c$ and where we
have only written down explicitly linear and quadratic terms in
$Q_a$. This action is invariant under the gauge transformations
$D_a{\longrightarrow}D_a$,
$Q_a{\longrightarrow}UQ_aU^{+}+U[D_a,U^{+}]$. Remark that the
background vector field $D_a$ appears here to play the same role
as the role played by ${L}_a$ in ordinary perturbation theory.
This means in particular that in order to fix the gauge in a
consistent way the gauge fixing term should be covariant with
respect to the background field. For example instead of the
Lorentz gauge $[L_a,Q_a]=0$ we impose here the covariant Lorentz
gauge $[D_a,Q_a]=0$. The gauge fixing term and the Faddeev-Popov
term are therefore given by

\begin{eqnarray}
 S_{g.f}+S_{gh}
& = &
-\frac{1}{2g^2}Tr_L\frac{[D_a,Q_a]^2}{\xi}+\frac{1}{g^2}Tr_Lb^{+}[D_a,[D_a,b]].
\label{ghost}
\end{eqnarray}
We will choose now for simplicity the gauge
${\xi}^{-1}=1+\frac{M^2}{|L|^2}$ which will cancel the $4$th term
in (\ref{action1}). This gauge becomes Feynman gauge ${\xi}=1$
in the limit $L{\longrightarrow}{\infty}$ and the Landau gauge
${\xi}=0$ in the limit $M{\longrightarrow}{\infty}$. Furthermore
we will assume that the background field $D_a$ satisfies the
classical equations of motion (\ref{eom}) and hence the $2$nd
term of (\ref{action1}) also vanishes. The partition function
becomes then
\begin{eqnarray}
Z_L[J]&=&e^{-S[D_a]-\frac{1}{g^2}Tr_LD_aJ_a}\det\big({{\cal D}^2}\big)\int
{\prod}_{a=1}^3\left[dQ_a\right]~e^{-\frac{1}{2g^2}Tr_LQ_a{\Omega}_{ab}Q_b+...},
\end{eqnarray}
where $Det\big({\cal D}^2\big)$ comes obviously from the integration over
the ghost field whereas the Laplacian ${\Omega}_{ab}$ is defined
by
\begin{eqnarray}
{\Omega}_{ab}&=&\alpha {\delta}_{ab}+ {\cal D
}_c^2{\delta}_{ab}+2{\cal
F}_{ab}+\frac{2M^2}{|L|^2}(D_c^2-|L|^2){\delta}_{ab}+\frac{4M^2}{|L|^2}D_aD_b.
\end{eqnarray}
In above the notation ${\cal D}_a$ and ${\cal F}_{ab}$ means that
the covariant derivative $D_a$ and the curvature $F_{ab}$ act by
commutators, i.e ${\cal D}_a(M)=[D_a,M]$, ${\cal
F}_{ab}(M)=[F_{ab},M]$ where $M{\in}Mat_{n(L+1)}$. Similarly
${\cal D}^2(f)=[D_a,[D_a,M]]$. Performing the Gaussian path
integral we obtain the one-loop effective action
\begin{eqnarray}
{\Gamma}[D_a]&=&S[D_a]+\frac{1}{2}Tr_3TR\log{\Omega}-TR\log{\cal
D}^2.\label{effective}
\end{eqnarray}
Note that the trace $TR$ appearing in this action is not the
trace $Tr_{L}$ over the indices of matrices but it is a trace over
$4$ indices corresponding to the left action and right action of
operators on matrices . $Tr_{3}$ means a trace associated with
$3-$dimensional rotations. As we will show the above result
holds also for $U(n)$ theories on the fuzzy sphere where only the
meaning of the different symbols becomes of course different.

\subsection{The $U(1)$ theory.}
In the following we will concentrate on 
the $U(1)$ model on the
fuzzy sphere with $M=\alpha=0$. It is not difficult to see that
the theory with $\alpha {\neq}0 $ involves adding the constant
$\alpha$ to the propagator and hence it is expected to have the
same limiting qualitative behaviour, whereas the theory with
$M{\neq}0$ will be studied in more detail in the next sections. We introduce the $U(1)$ gauge field by writing $D_a=L_a+A_a$. Although we are not going to use the Feynman rules explicitly it will be instructive to write them down here for completeness. They are extracted from the last two terms of (\ref{effective}) or more precisely from the action
\begin{eqnarray}
-\frac{1}{2g^2}Tr_LQ_a{\Omega}_{ab}Q_b+\frac{1}{g^2}Tr_Lb^+{\cal D}^2b.\label{I}
\end{eqnarray}
The propagators of the fluctuation fields $Q_a$ and the ghost fields $b^{+}$ and $b$ are found to be given by the inverse of the Laplacian ${\cal L}^2$ (see Figure $1a$ and Figure $1b$ respectively). Indeed it is not difficult to see from (\ref{I}) that the quadratic actions reads 
\begin{eqnarray}
-\frac{1}{2g^2}Tr_LQ_a{\cal L}^2Q_a+\frac{1}{g^2}Tr_Lb^{+}{\cal L}^2b.
\end{eqnarray}   
There are also cubic vertices involving the fluctuation field $Q_a$ , the gauge field $A_a$ and the ghost fields $b$ and $b^{+}$. The $AQQ$ vertex , the ${\cal F}^{0}QQ$ and the $b^{+}bA$ vertex are given respectively by the operators
\begin{eqnarray}
&&-\frac{1}{g^2}Tr_LQ_a\left(\frac{{\cal L}{\cal A}+{\cal A}{\cal L}}{2}\right)Q_a\nonumber\\
&&-\frac{1}{g^2}Tr_LQ_a{\cal F}^{(0)}_{ab}Q_b\nonumber\\
&&+\frac{1}{g^2}Tr_Lb^{+}({\cal L}{\cal A}+{\cal A}{\cal L})b. 
\end{eqnarray}
The relevant Feynman graphs are given in Figures $2a$, $2b$ and $2c$. Let us remark here that there is no coupling between the ghost fields and the curvature. The quartic vertices are given on the other hand by the interactions $AAQQ$ and $AAb^{+}b$. Explicitly they are given by the following terms in the action 
\begin{eqnarray}
&&-\frac{1}{g^2}Tr_LQ_a\left(\frac{[A_c,[A_c,~\cdot]]{\delta}_{ab}+2[[A_a,A_b],~\cdot]}{2}\right)Q_b\nonumber\\
&&+\frac{1}{g^2}Tr_Lb^{+}{\cal A}^2b. 
\end{eqnarray}
The corresponding Feynman graphs are given in  $3a$, $3b$.

The first term in (\ref{effective}) gives the full tree-level action of the gauge field $A_a$. This reads
\begin{eqnarray}
S[D_a]=-\frac{1}{4g^2}Tr_LF_{ab}^2-\frac{i}{2g^2}{\epsilon}_{abc}Tr_L\big[\frac{1}{2}F_{ab}^{(0)}A_c+\frac{1}{3}[A_a,A_b]A_c\big]=S_2+S_3+S_4,\label{llll}
\end{eqnarray}
where $F_{ab}=F_{ab}^{(0)}+[A_a,A_b]$, $F_{ab}^{(0)}={\cal
L}_aA_b-{\cal L}_bA_a-i{\epsilon}_{abc}A_c$ and $S_2$, $S_3$ and
$S_4$ are the quadratic, cubic and quartic actions respectively of the gauge field $A_a$.
In particular the quadratic action $S_2$ reads
\begin{eqnarray}
S_2=-\frac{1}{2g^2}Tr_L[L_a,A_b]^2+\frac{1}{2g^2}Tr_L[L_a,A_a]^2.\label{lll}
\end{eqnarray}
We will apply directly the result (\ref{effective}) to find quantum
corrections to this quadratic action. This will of course
capture all quantum corrections to the vacuum polarization tensor
as well as tadpole corrections. The quadratic effective action
is given by
\begin{eqnarray}
{\Gamma}_2  &=&S_2+\frac{1}{2}TR \log{\cal
L}^2+\frac{1}{2}{TR}\left(\Delta^{(1)}+\Delta^{(2)}-\frac{1}{2}(\Delta^{(1)})^2\right)-\frac{1}{4}{TR}(\Delta^{(j)})^2_{aa}
\label{ex}
\end{eqnarray}
where we have only kept constant, linear and quadratic terms in
the gauge field $A_a$ and where ${\Delta}^{(1)}$,
${\Delta}^{(2)}$ and ${\Delta}^{(j)}$ are defined by

\begin{eqnarray}
\Delta^{(1)}  = \frac{1}{{\cal L}^2}\big({\cal L}{\cal A}+{\cal
A}{\cal L}\big)~,~\Delta^{(2)}  =  \frac{1}{{\cal L}^2}{\cal
A}^2~,~{\Delta}^{(j)}_{ab}=\frac{2}{{\cal L}^2}{\cal
F}^{(0)}_{ab}.
\end{eqnarray}
In the appendices we will give the detail of the computation and we summarize the corresponding results below. In this computation  we use extensively the following Green's function
\begin{eqnarray}
\left(\frac{1}{\mathcal{L}^2}\right)^{AB,CD}=\frac{1}{L+1}\sum_{lm}\frac{1}{l(l+1)}\hat{Y}_{lm}^{AB}(\hat{Y}_{lm}^{+})^{DC}.
\end{eqnarray}
Remark that since the propagator $({\cal L}^2)^{-1}$ acts on the algebra of matrices it carries 4 indices. We will also need to use  the following identities
\begin{eqnarray}
TR\big(M\frac{1}{{\cal L}^2}{\cal
O}\big)=\sum_{lm}\frac{1}{l(l+1)}Tr_L\hat{Y}_{lm}^{+}{\cal
O}\big(M\hat{Y}_{lm}\big)
\label{identity1}
\end{eqnarray}
and
\begin{eqnarray}
TR\big(M\frac{1}{{\cal L}^2}{\cal O}\frac{1}{{\cal L}^2}N\big)
=\sum_{lm}\sum_{kn}\frac{1}{l(l+1)}\frac{1}{k(k+1)}Tr_L\left[\hat{Y}_{lm}\hat{Y}_{kn}^{+}NM\right]
Tr_L \left[\hat{Y}_{lm}^{+}{\cal O}(\hat{Y}_{kn})\right].
\label{identity2}
\end{eqnarray}
In above $M$ and $N$ are two arbitrary matrices and ${\cal O}$ is
some operator acting on the space of these matrices.

\paragraph{Tadpole contribution.}  Quantum correction to the tree-level linear term which is
identically zero, i.e $S_1=0$, is given by the combination of the two
tadpole diagrams of Figure $4$. These diagrams are also equal to
the third term in  expansion (\ref{ex}), viz $ {\Gamma}_1 =
\frac{1}{2}TR\Delta^{(1)}$. By writing
$A_a=\sum_{pn}A_a(pn)\hat{Y}_{pn}$ and $\Gamma_{1}$ as
\begin{equation}
\Gamma_{1}=\frac{1}{2}\sum_{pn}\frac{Tr_{L}\left[L_{a},\hat{Y}^{\dagger}_{pn}\right]\left[A_{a},\hat{Y}_{pn}\right]}{p\left(p+1\right)}
\end{equation} 
we can compute, using the
different identities of \cite{VKM},  the action ${\Gamma}_1$.
One finds the result
\begin{eqnarray}
{\Gamma}_1&=&\frac{4}{\sqrt{3}}|L|A_{-\mu}(1-\mu)=4|L|Tr_Lx_aA_a.
\end{eqnarray}
Now  we can use the definition of the normal scalar
field on the fuzzy sphere given by
$\phi=\frac{1}{2}(x_aA_a+A_ax_a+\frac{A_a^2}{|L|})$ to rewrite this expression in the form
\begin{eqnarray}
{\Gamma}_1&=&4|L|Tr_L\phi
-2Tr_LA_a^2.\label{5.21}
\end{eqnarray}
This identity is exact and as it turns out this separation is crucial in establishing covariance of the  $U(1)$ theory in the fuzzy setting in the sense of equation (\ref{eff1}) below.

\paragraph{Vacuum polarization tensor.}

{\it The $4-$vertex contribution} to the vacuum polarization tensor is
given by the diagrams of Figure $5$ which are also equal to the
$4$th term in the expansion (\ref{ex}), i.e $ {\Gamma}_2^{(4)} =
\frac{1}{2}{TR}\Delta^{(2)}$. This can be put in the form
\begin{eqnarray}
{\Gamma}_2^{(4)} =  -\frac{1}{2}\sum_{l_1m_1}\frac{Tr_L[A_a,\hat{Y}_{l_1m_1}^{\dagger}][A_a,\hat{Y}_{l_1m_1}]}{l_1(l_1+1)}
\end{eqnarray}
After a long calculation we get
the explicit answer
\begin{eqnarray}
{\Gamma}_2^{(4)}=Tr_LA_a{\cal L}^2{\Delta}_4A_a,\label{5.26}
\end{eqnarray}
with the conservation law that $p_1+l_1+l_2$ must be an odd
number and where  the eigenvalues of the operator
${\Delta}_4{\equiv}{\Delta}_4({\cal L}^2)$ on
the eigenvectors $\hat{Y}_{p_1n_1}$ are given by
\begin{eqnarray}
{\Delta}_4(p_1)=\sum_{l_1,l_2}\frac{2l_1+1}{l_1(l_1+1)}\frac{2l_2+1}{l_2(l_2+1)}(1-(-1)^{l_1+l_2+p_1})(L+1)\left\{\begin{array}{ccc}
        p_1 & l_1 & l_2 \\
    \frac{L}{2} & \frac{L}{2} & \frac{L}{2}
    \end{array} \right\}^2\frac{l_2(l_2+1)}{p_1(p_1+1)}.\label{5.27}
\end{eqnarray}

{\it The $3-$vertex contribution} comes from three different diagrams.
The contribution of the ${\cal F}$ term is given by the diagram of Figure $6b$ and it corresponds to the last term in expansion
(\ref{ex}), namely $
{\Gamma}_2^{(3F)}=-\frac{1}{4}TR({\Delta}^{(j)})^2_{aa}$ whereas
the $5$th term in  expansion (\ref{ex}), i.e $
{\Gamma}_2^{(3A)}=-\frac{1}{4}TR({\Delta}^{(1)})^2$, corresponds
to the combination of the diagrams displayed in Figure $6a$. The
diagram of Figure $6b$  can be represented by

\begin{eqnarray}
{\Gamma}_2^{(3F)}=\sum_{k_1m_1}\sum_{k_2m_2}\frac{Tr_L \left[ F_{ab}^{(0)}[\hat{Y}_{k_2m_2},\hat{Y}_{k_1m_1}^{\dagger}]\right]Tr_L \left[F_{ab}^{(0)}[\hat{Y}_{k_1m_1},\hat{Y}^{\dagger}_{k_2m_2}]\right]
}{k_1\left(k_1+1\right)k_2\left(k_2+1\right)}
\end{eqnarray}
For this diagram a long calculation yields  the explicit
result
\begin{eqnarray}
{\Gamma}_2^{(3F)}=Tr_LF_{ab}^{(0)}{\Delta}_FF_{ab}^{(0)},\label{5.31}
\end{eqnarray}
where the operator ${\Delta}_F{\equiv}{\Delta}_F({\cal L}^2)$
is defined by its spectrum
\begin{eqnarray}
{\Delta}_F(p_1)&=&2\sum_{l_1,l_2}\frac{2l_1+1}{l_1(l_1+1)}\frac{2l_2+1}{l_2(l_2+1)}(1-(-1)^{l_1+l_2+p_1})(L+1)\left\{\begin{array}{ccc}
        l_1 & l_2 & p_1 \\
    \frac{L}{2} & \frac{L}{2} & \frac{L}{2} \end{array}\right\}^2.\label{5.32}
\end{eqnarray}
The quantum action ${\Gamma}_2^{(4)}$ given by (\ref{5.26}) is
clearly a correction to the first term of (\ref{lll}) whereas the
action ${\Gamma}_2^{(3F)}$ given by (\ref{5.31}) contains a
correction to both terms in (\ref{lll}) plus a mass term.

Similarly the diagrams of Figure $6a$ admit the representation

\begin{eqnarray}
 \Gamma_2^{(3(A))}   = & -&\frac{1}{2}\sum_{l_1m_1}\sum_{l_2m_2}\frac{{Tr}_L[L_a,\hat{Y}_{l_1m_1}][A_a,\hat{Y}_{l_2m_2}^{+}]
           {Tr}_L[L_b,\hat{Y}_{l_2m_2}][A_b,\hat{Y}_{l_1m_1}^{+}]}{l_1(l_1+1)l_2(l_2+1)}\nonumber\\
&  - & \frac{1}{2}\sum_{l_1m_1}\sum_{l_2m_2}\frac{{Tr}_L[L_a,\hat{Y}_{l_1m_1}][A_a,\hat{Y}_{l_2m_2}^{+}]
           {Tr}_L [L_b,\hat{Y}_{l_1m_1}^{+}][A_b,\hat{Y}_{l_2m_2}]}{l_1(l_1+1)l_2(l_2+1)}
\end{eqnarray}
Explicitly we find for these diagrams the result
\begin{eqnarray}
\Gamma_2^{(3A)}&=&-2\sum_{p_1n_1}\sum_{p_2n_2}A_{-\mu}(p_1n_1)A_{-\nu}(p_2n_2)(-1)^{n_1+\nu}\sum_{l_1,l_2}\frac{2l_1+1}{l_1(l_1+1)}\frac{2l_2+1}{l_2(l_2+1)}(L+1)\nonumber\\
&{\times}&\left\{\begin{array}{ccc}
        p_1 & l_1 & l_2 \\
    \frac{L}{2} & \frac{L}{2} & \frac{L}{2} \end{array}\right\} \left\{\begin{array}{ccc}
        p_2 & l_1 & l_2 \\
    \frac{L}{2} & \frac{L}{2} & \frac{L}{2}
    \end{array}\right\}f^{(A)}(lpn;\mu,\nu).\label{lpn}
\end{eqnarray}
In this case we must also have the conservation laws
$l_1+l_2+p_1={\rm odd}~{\rm number}$, $l_1+l_2+p_2={\rm
odd}~{\rm number}$ which means in particular that $p_1+p_2$ can
only be an even number. The function $f^{(A)}$ is of the form
$f^{A}=f^{A_1}+f^{A_2}$ where in particular
\begin{eqnarray}
f^{A_1}=-\frac{C_{p_1n_11\mu}^{p_1m}C_{p_2n_21\nu}^{p_1-m}{\delta}_{p_1p_2}}
{2 p_1(p_1+1)}\left[l_2(l_2+1)-l_1(l_1+1)\right]\left[l_2(l_2+1)-l_1(l_1+1)-p_1(p_1+1)\right].
\end{eqnarray}
We can see therefore that the contribution $ \Gamma_2^{(3A)}$
splits into two parts, a canonical gauge contribution $
\Gamma_2^{(3A_1)}$ plus a non-trivial part $ \Gamma_2^{(3A_2)}$
corresponding to whether $f$ is equal to $f^{A_1}$ or $f^{A_2}$
respectively. The canonical gauge part is explicitly given by
the expression
\begin{eqnarray}
\Gamma_2^{(3A_1)}=-Tr_LA_a{\cal L}_a{\Delta}_3{\cal
L}_bA_b,\label{5.38}
\end{eqnarray}
where again the operator ${\Delta}_3{\equiv}{\Delta}_3({\cal L}^2)$ is defined by its spectrum
\begin{eqnarray}
{\Delta}_3(p_1)&=&\sum_{l_1,l_2}\frac{2l_1+1}{l_1(l_1+1)}\frac{2l_2+1}{l_2(l_2+1)}(1-(-1)^{l_1+l_2+p_1})(L+1)\left\{\begin{array}{ccc}
        p_1 & l_1 & l_2 \\
    \frac{L}{2} & \frac{L}{2} & \frac{L}{2} \end{array}\right\}^2\nonumber\\
&{\times}&\frac{l_2(l_2+1)}{p_1^2(p_1+1)^2}\big(l_2(l_2+1)-l_1(l_1+1)\big).\label{5.37}
\end{eqnarray}
This is clearly a correction to the kinetic term in the tree-level
action (\ref{lll}). We remark that so far all quantum corrections
to the vacuum polarization tensor given by equations (\ref{5.26}),
 (\ref{5.31}) and (\ref{5.38}) are written in terms of the
operator
\begin{eqnarray}
{\Delta}(p_1,p_2)&=&\sum_{l_1l_2}\frac{2l_1+1}{l_1(l_1+1)}\frac{2l_2+1}{l_2(l_2+1)}(L+1)\left\{\begin{array}{ccc}
        p_1 & l_1 & l_2 \\
    \frac{L}{2} & \frac{L}{2} & \frac{L}{2} \end{array}\right\} \left\{\begin{array}{ccc}
        p_2 & l_1 & l_2 \\
    \frac{L}{2} & \frac{L}{2} & \frac{L}{2}
    \end{array}\right\}X(l_1,l_2,p_1,p_2),\label{5.47}\nonumber\\
\end{eqnarray}
where $X$ , for all these actions (\ref{5.26}), (\ref{5.31}) and
(\ref{5.38}, is of the form
$X(l_1,l_2,p_1,p_2)=\delta_{p_1p_2}\bar{X}(l_1,l_2,p_1)$ and
where the sums are always over $l_1$ and $l_2$ such that
$l_1+l_2+p_1$ is an odd number.

The last correction to the vacuum polarization tensor is given by
equation (\ref{lpn}) where the function $f^{(A)}$ is replaced by
$f^{(A_2)}$. This leads to a more complicated contribution which
we can write in the form

\begin{eqnarray}
\Gamma_{2}^{(3A_2)}&=&\sum_{p_1n_1}\sum_{p_2n_2}A_{-\mu}(p_1n_1)A_{-\nu}(p_2n_2)(-1)^{n_1+\nu}\bigg[C_{p_1n_11\mu}^{p_1-1m}C_{p_2n_21\nu}^{p_1-1-m}\big({\Lambda}^{(-)}(p_1,p_2)+ {\Sigma}^{(-)}(p_1,p_2)\big)\nonumber\\
&+&C_{p_1n_11\mu}^{p_1+1m}C_{p_2n_21\nu}^{p_1+1-m}\big({\Lambda}^{(+)}(p_1,p_2)+{\Sigma}^{(+)}(p_1,p_2)\big)\bigg].\label{46}
\end{eqnarray}
The functions ${\Lambda}^{(\pm)}(p_1,p_2)$ are of the form
(\ref{5.47}) with some $X$ such that
{\footnotesize{$\Lambda^{\left(\pm\right)}\left(p_1,p_2\right)=\delta_{p_{1},p_{2}}\bar{\Lambda}^{\left(\pm\right)}\left(p_1\right)$}}
whereas the functions ${\Sigma}^{(\pm)}(p_1,p_2)$ are the form
(\ref{5.47}) but with $X$'s such that
${\Sigma}^{(\pm)}(p_1,p_2)={\delta}_{p_1{\pm}2,p_2}\bar{\Sigma}^{(\pm)}(p_1)$.
The explicit expressions of the corresponding $X$'s is not
important for the purpose of this section and thus we simply skip
writing them down here  . They are given in appendix $C$ . Finally we can see by inspection that the
Clebsch-Gordan coefficients appearing in the action $
\Gamma_{2}^{(3A_2)}$  are exactly those which appear in  the
scalar mass term in the action . Indeed we can compute
\begin{eqnarray}
& &\frac{1}{4}Tr_L[x_a,A_a]_{+}^2=\sum_{p_1n_1}\sum_{p_2n_2}A_{-\mu}(p_1n_1)A_{-\nu}(p_2n_2)(-1)^{n_1+\nu}\bigg[C_{p_1n_11\mu}^{p_1-1m}C_{p_2n_21\nu}^{p_1-1-m}\nonumber\\
&\times&\big({\lambda}^{(-)}(p_1,p_2)+ {\sigma}^{(-)}(p_1,p_2)\big)
+C_{p_1n_11\mu}^{p_1+1m}C_{p_2n_21\nu}^{p_1+1-m}\big({\lambda}^{(+)}(p_1,p_2)+{\sigma}^{(+)}(p_1,p_2)\big)\bigg],\nonumber\\
\label{47}
\end{eqnarray}
where ${\lambda}^{(\pm)}(p_1,p_2)$ and ${\sigma}^{(\pm)}(p_1,p_2)$
are some other functions which are such that
${\lambda}^{(\pm)}(p_1,p_2)={\delta}_{p_1,p_2}\bar{\lambda}^{(\pm)}(p_1)$,
${\sigma}^{(\pm)}(p_1,p_2)={\delta}_{p_1{\pm}2,p_2}\bar{\sigma}^{(\pm)}(p_1)$.
These functions are classical and hence they are not loops of the
form (\ref{5.47}) as it must be obvious (see appendix $C$). By
comparing (\ref{46}) and (\ref{47}) we can immediately deduce
that the action ${\Gamma}_2^{(3A_2)}$ in position space must
involve anticommutators of $x_a$ and $A_a$ instead of commutators
and hence it is of a scalar-like type. As we will show this
action will still contain (in the limit) terms which describe
non-local interactions between the scalar and the gauge fields.
In configuration it reads

\begin{eqnarray}
 \Gamma_{2}^{(3A_2)}&=&\sum_{ij}
Tr_L[{\nabla}_i(A_a),x_a]_{+}{\Delta}_{ij}\big([{\nabla}_j(A_b),x_b]_{+}\big).
\label{5.39}
\end{eqnarray}
The operators ${\Delta}_{ij}$ are some combinations of the
operators  ${\Lambda}^{(\pm)}$ and ${\Sigma}^{(\pm)}$ whereas the
operators ${\nabla}_i$ are sign operators of the form
${\nabla}_i=(-1)^{{\alpha}_i\hat{N}}$, $\hat{N}\hat{Y}_{lm}=l\hat{Y}_{lm}$. Again the corresponding explicit expressions are found in appendix $C$.

\subsection{Gauge covariance on $S^2_L$.}
  By putting together equations (\ref{5.21}), (\ref{5.26}),
(\ref{5.31}), (\ref{5.38}) and (\ref{5.39}) we obtain the full
quadratic $U(1)$ effective action on ${\bf S}^2_L$, namely
\begin{eqnarray}
{\Gamma}_{2}&=&S_2+\frac{1}{2}TRlog{\cal
L}^2+4|L|Tr_L\phi+Tr_LA_a\big({\cal
L}^2{\Delta}_4-2\big)A_a-Tr_LA_a{\cal L}_a{\Delta}_3{\cal
L}_bA_b\nonumber\\&+&Tr_LF_{ab}^{(0)}{\Delta}_FF_{ab}^{(0)}
+{\Gamma}_2^{(3A_2)}.\label{eff}
\end{eqnarray}
In  is rather clear that the first $3$ terms are gauge invariant.
Naturally we also expect that the $4$th and $5$th terms in
(\ref{eff}) to become gauge invariant in the continuum limit. To
check this property explicitly we rewrite these two terms as
follows
\begin{eqnarray}
Tr_LA_a\big({\cal L}^2{\Delta}_4-2\big)A_a-Tr_LA_a{\cal L}_a{\Delta}_3{\cal L}_bA_b&=&-\frac{1}{2}Tr_LF_{ab}^{(0)}{\Delta}_3F_{ab}^{(0)}\nonumber\\
&-&\frac{i}{2}{\epsilon}_{abc}Tr_LF_{ab}^{(0)}\big({\Delta}_3+{\cal L}^2({\Delta}_3-{\Delta}_4)+2\big)A_c\nonumber\\
&+&i{\epsilon}_{abc}Tr_L{\cal L}_aA_b\big({\cal L}^2({\Delta}_3-{\Delta}_4)+2\big)A_c.
\end{eqnarray}
The first two terms in this expression are now exactly gauge invariant in the continuum limit whereas the third term it can not be gauge invariant unless it vanishes identically. We expect therefore by the requirement of gauge invariance alone that we have the asymptotic behaviour $
{\cal L}^2({\Delta}_3-{\Delta}_4)+2{\longrightarrow}0~,~L{\longrightarrow}{\infty}$.
As it turns out this statement is true for all finite values of $L$, in other words we have in fact the identity
\begin{eqnarray}
{\cal L}^2({\Delta}_3-{\Delta}_4)+2=0.\label{5.42}
\end{eqnarray}
Indeed we have from (\ref{5.27}) and (\ref{5.37}) the difference

\begin{eqnarray}
{\Delta}_3(p_1)-{\Delta}_4(p_1)&=&\frac{(L+1)}{p_1^2(p_1+1)^2}\sum_{l_1,l_2}\frac{2l_1+1}{l_1(l_1+1)}\frac{2l_2+1}{l_2(l_2+1)}(1-(-1)^{l_1+l_2+p_1})\left\{\begin{array}{ccc}
        p_1 & l_1 & l_2 \\
    \frac{L}{2} & \frac{L}{2} & \frac{L}{2} \end{array}\right\}^2 \nonumber\\
&{\times}&l_2(l_2+1)\left[l_2(l_2+1)-l_1(l_1+1)-p_1(p_1+1)\right]\nonumber\\
&=&-\frac{2(L+1)}{p_1^2(p_1+1)^2}\sqrt{p_1(p_1+1)(2p_1+1)}\sum_{l_1}\frac{2l_1+1}{l_1(l_1+1)}\sqrt{l_1(l_1+1)(2l_1+1)}\nonumber\\
&{\times}&\sum_{l_2}(2l_2+1)(1-(-1)^{l_1+l_2+p_1})\left\{\begin{array}{ccc}
        p_1 & l_1 & l_2 \\
    \frac{L}{2} & \frac{L}{2} & \frac{L}{2} \end{array}\right\}^2
\left\{\begin{array}{ccc}
        l_2 & l_1 & p_1 \\
    1 & p_1 & l_1 \end{array}\right\}\nonumber\\
&=&-\frac{2\left(L+1\right)}{p_1^2(p_1+1)^2}\sqrt{p_1(p_1+1)(2p_1+1)}\sum_{l_1}\frac{2l_1+1}{l_1(l_1+1)}\sqrt{l_1(l_1+1)(2l_1+1)}\nonumber\\
&{\times}&\bigg[\left\{\begin{array}{ccc}
        p_1 & p_1 & 1 \\
    \frac{L}{2} & \frac{L}{2} & l_1 \\
\frac{L}{2} & \frac{L}{2} & l_1 \end{array}\right\}-(-1)^{l_1+p_1+1}\left\{\begin{array}{ccc}
        \frac{L}{2} & \frac{L}{2} & 1 \\
    p_1 & p_1 & \frac{L}{2} \end{array}\right\}\left\{\begin{array}{ccc}
        \frac{L}{2} & \frac{L}{2} & 1 \\
    l_1 & l_1 & \frac{L}{2} \end{array}\right\}\bigg],
\end{eqnarray}
where we have used in the second line the identity
\begin{eqnarray}
\left\{\begin{array}{ccc}
        l_2 & l_1 & p_1 \\
    1 & p_1 & l_1 \end{array}\right\}=-\frac{1}{2}\frac{l_2(l_2+1)-l_1(l_1+1)-p_1(p_1+1)}{\sqrt{l_1(l_1+1)(2l_1+1)p_1(2p_1+1)(p_1+1)}},
\end{eqnarray}
then performed the sum over $l_2$ by using equations $(5)$ and $(6)$ on page $305$ of \cite{VKM}. Similarly we can use the explicit expressions of the resulting $9j$ and $6j$ symbols and then do the sum over $l_1$ to obtain the final result

\begin{eqnarray}
{\Delta}_3(p_1)-{\Delta}_4(p_1)=-\frac{2}{p_1(p_1+1)}.\label{5.45}
\end{eqnarray}
This is exactly equation (\ref{5.42}). Thus the quadratic effective action on the fuzzy sphere reads
\begin{eqnarray}
{\Gamma}_{2}&=&S+\frac{1}{2}\sum_{l=1}^L(2l+1)Logl(l+1)+4|L|Tr_L\phi+Tr_LF_{ab}^{(0)}\big({\Delta}_F-\frac{1}{2}{\Delta}_3\big)F_{ab}^{(0)}\nonumber\\
&-&\frac{i}{2}{\epsilon}_{abc}Tr_LF_{ab}^{(0)}{\Delta}_3A_c
+{\Gamma}_2^{(3A_2)}.\label{eff1}
\end{eqnarray}
From this last expression it is now obvious that the $4th$ and $5th$ terms in this effective action become gauge invariant in the continuum limit. It remains now to establish gauge invariance of the last term
${\Gamma}_2^{(3A_2)}$. To see this let us recall that $U(1)$
gauge transformations on the continuum sphere act as follows
$A_a{\longrightarrow}A_a^{\Lambda}=A_a-i{\cal L}_a(\Lambda)$. Let
us also observe that in equation (\ref{5.39}) the operators
${\nabla}_i$ depend only on ${\cal L}^2$ and since $[{\cal
L}^2,{\cal L}_a]=0$ we can show that ${\nabla}_i(A_a^{\Lambda})=
{\nabla}_i(A_a)-i{\cal L}_a({\nabla}_i({\Lambda}))$.
Hence in the limit we will have the identity
$x_a{\nabla}_i(A_a^{\Lambda})=x_a{\nabla}_i(A_a)$ and as a
consequence the action ${\Gamma}_2^{(3A_2)}$ is gauge invariant
as expected. 

This in fact establishes gauge invariance  of the
above quadratic effective action in the continuum large $L$ limit. However in order to show gauge invariance of the whole model for finite $L$ we need also to compute the quantum corrections to the cubic and quartic vertices.

\section{The continuum limit and the UV-IR mixing}

\subsection{The UV-IR mixing.} 

The criterion for the existence of a UV-IR mixing phenomena on the fuzzy sphere is defined by the requirement that the fuzzy quantum effective action does not approach the quantum effective action on the commutative sphere. For a $U(1)$ theory on ordinary $S^2$ the action (\ref{higgs}) is quadratic 
in the fields $a_{a}$ and $\phi$ and thus the quantum corrections are trivial, i.e the effective action on $S^2$ is essentially equal to the classical action (\ref{higgs}). On the other hand the quantum corrections on the fuzzy sphere yield the action ${\Gamma}_2$. So to show the existence of a UV-IR mixing phenomena in this model we need
only to check that some (or all) of the operators in the above
effective action (\ref{eff1}) do not vanish in the continuum
limit. In other words  we need to show that ${\Gamma}_2$ does not tend to $S$ in the limit. Since each  term of the action (\ref{eff1}) will be gauge
invariant in the continuum limit it is safe  to study  the
continuum limit of the individual operators ${\Delta}_3$,
${\Delta}_4$, ${\Delta}_{F}$ and those appearing in
${\Gamma}_2^{(3A_2)}$.

In this section we concentrate only on the operators $\Delta_3$, 
$\Delta_4$ and $\Delta_{F}$. We go back to equation
(\ref{5.27}) and rewrite the loop ${\Delta}_4(p_1)$ in the form
\begin{eqnarray}
{\Delta}_4(p_1)=\frac{\left(L+1\right)}{p_1(p_1+1)}\sum_{l_1}\frac{2l_1+1}{l_1(l_1+1)}\sum_{l_2}(2l_2+1)(1-(-1)^{l_1+l_2+p_1})\left\{\begin{array}{ccc}
        p_1 & l_1 & l_2 \\
    \frac{L}{2} & \frac{L}{2} & \frac{L}{2}
    \end{array} \right\}^2\frac{l_2(l_2+1)}{p_1(p_1+1)}.\nonumber\\
\end{eqnarray}
We can immediately notice that it is  possible in fact to do the
sum over $l_2$ using the results
\begin{eqnarray}
 \sum_{l_2}(2l_2+1)\left\{\begin{array}{ccc}
        p_1 & l_1 & l_2 \\
    \frac{L}{2} & \frac{L}{2} & \frac{L}{2}
    \end{array} \right\}^2=\frac{1}{L+1}
\end{eqnarray}
and
\begin{eqnarray}
\sum_{l_2}(2l_2+1)(-1)^{l_2}\left\{\begin{array}{ccc}
        p_1 & l_1 & l_2 \\
    \frac{L}{2} & \frac{L}{2} & \frac{L}{2}
    \end{array} \right\}^2=(-1)^{L}\left\{\begin{array}{ccc}
        p_1 & \frac{L}{2} & \frac{L}{2} \\
    l_1 & \frac{L}{2} & \frac{L}{2}
    \end{array} \right\}.
\end{eqnarray}
${\Delta}_4$ takes therefore the simple form

\begin{equation}
{\Delta}_4(p_1)=\frac{1}{p_1(p_1+1)}\sum_l\frac{2l+1}{l(l+1)}\left[1-(L+1)(-1)^{l_1+p_1+L}\left\{\begin{array}{ccc}
        p_1 & \frac{L}{2} & \frac{L}{2} \\
    l_1 & \frac{L}{2} & \frac{L}{2}
    \end{array} \right\}\right].
\end{equation}
As it turns out 
we can use in the large $L$ limit the same approximation used in \cite{uvir}, namely
\begin{eqnarray}
\left\{ \begin{array}{ccc}
        p_1 & \frac{L}{2} & \frac{L}{2} \\
    l_1 & \frac{L}{2} & \frac{L}{2}
    \end{array} \right\}{\simeq}\frac{(-1)^{L+p_{1}+l_1}}{L}P_{p_1}(1-\frac{2l_1^2}{L^2}),
\end{eqnarray}
where  $P_{p}$ are the Legendre polynomials which are defined by the generating function 
\begin{eqnarray}
\frac{1}{\sqrt{1-2tx+t^2}}=\sum_{p=0}^{\infty}P_{p}(x)t^p.
\label{gene}
\end{eqnarray}
It is quite straightforward to conclude that in the limit we must
have
\begin{eqnarray}
{\Delta}_4(p_1)=-\frac{1}{p_1(p_1+1)}\int_{-1}^{1}\frac{dx}{1-x}\big[P_{p_1}(x)-1\big]{\equiv}-\frac{h(p_1)}{p_1(p_1+1)},
\end{eqnarray}
while by equation (\ref{5.45}) we must also have
\begin{eqnarray}
{\Delta}_3(p_1)=-\frac{h(p_1)+2}{p_1(p_1+1)}.\label{introduction}
\end{eqnarray}
The number $h(p_1)$ can be evaluated using the following trick.
We regularize the integral by replacing the upper bound by
$1-{\delta}$. By using (\ref{gene})
one have the following result
\begin{eqnarray}
\sum_{p=1}^{\infty}h(p)t^p&=&\frac{2}{1-t}\int_{1+\bar{\delta}}^{\frac{1+t}{1-t}}\frac{d{\alpha}}{{\alpha}^2-1}+\frac{1}{1-t}ln\frac{\delta}{2}=\frac{1}{1-t}ln\frac{t{\delta}}{\bar{\delta}}=\frac{2}{1-t}ln(1-t)~,~\bar{\delta}=\frac{t}{(1-t)^2}{\delta}.\nonumber\\
\end{eqnarray}
We can further write the above equation as follows
\begin{eqnarray}
\sum_{p=1}^{\infty}h(p)t^p&=&-2\sum_{p=0}^{\infty}\sum_{k=1}^{\infty}\frac{t^{p+k}}{k}=-2\sum_{p=1}^{\infty}\sum_{k=1}^{p}\frac{t^p}{k},\nonumber
\end{eqnarray}
and thus
$h(p_1)=-2\sum_{l=1}^{p_1}\frac{1}{l}$. Hence one obtains
\begin{eqnarray}
{\Delta}_4(p_1)=\frac{2\sum_{l=1}^{p_1}\frac{1}{l}}{p_1(p_1+1)}~,~
{\Delta}_3(p_1)=\frac{2\sum_{l=2}^{p_1}\frac{1}{l}}{p_1(p_1+1)}.
\end{eqnarray}
Putting the above results together we obtain in the continuum the
effective action
\begin{eqnarray}
{\Gamma}_{2}&=&S_2+\frac{1}{2}\sum_{l=1}^L\left(2l+1\right)\log l(l+1) -\frac{1}{2}\int
\frac{d{\Omega}}{4{\pi}}F_{ab}^{(0)}{\Delta}_3F_{ab}^{(0)}-\frac{i}{2}{\epsilon}_{abc}\int
\frac{d{\Omega}}{4{\pi}}F_{ab}^{(0)}{\Delta}_3A_c\nonumber\\
&+&4|L|\int\frac{d{\Omega}}{4{\pi}}\phi +{\Gamma}_2^{(3A_2)}.\label{eff2}
\end{eqnarray}
In above we have also used the result that the operator ${\Delta}_F$ approaches zero in the continuum limit (see below for the proof). The first correction to the classical action $S_2$  is just an
infinite constant. The $2$nd and $3$th of the effective action (\ref{eff2}) 
corrections  clearly give rise to a non-trivial quantum
contribution to the $U(1)$ action on ${\bf S}^2$ due to the fuzzy
sphere, i.e the $U(1)$ theory on ${\bf S}^2$ obtained as a limit
of a $U(1)$ theory on ${\bf S}^2_L$ is not a simple Gaussian
theory. These two terms reflect the existence of a gauge
invariant UV-IR mixing on the fuzzy sphere which survives the
continuum limit. The $4$th correction is only relevant for the scalar sector of the theory
and thus it does not lead to any noncommutative anomaly or UV-IR
mixing in the $2-$dimensional gauge sector. However we still have to study  the continuum
limit of the last term ${\Gamma}_2^{(3A_2)}$ in some more detail.

\subsection{The continuum limit.} 

In this subsection we will study in detail the continuum limit of
the effective action ${\Gamma}_2^{(3A_2)}$. As we have said
before the operators ${\Delta}_{ij}$ appearing in (\ref{5.39})
are of the form (\ref{5.47}) with some complicated $X$'s. It is
expected that the loop (\ref{5.47}) will be dominated in the large
$L$ limit by the UV region of the internal momenta $l_1$ and
$l_2$, i.e we should be able to assume for all practical
purposes that $l_1$ and $l_2$ are much larger compared to $1$.
To be precise let us split the sum $\sum_{l_1l_2}$ as follows
$\sum_{l_1=0}^{L}\sum_{l_2=|l_1-p_1|}^{l_1+p_1}=\sum_{l_1=0}^{\Lambda}\sum_{l_2=|l_1-p_1|}^{l_1+p_1}+\sum_{l_1=\Lambda}^{L}\sum_{l_2=|l_1-p_1|}^{l_1+p_1}$
where $\Lambda$ is an intermediary cut-off which is such that
$1<<\Lambda<<L$. Under the first sum, and provided we
concentrate on the regime where the external momenta $p_1$ and
$p_2$ are much less than the cut-off $\Lambda$, we can treat the
internal momenta $l_1$, $l_2$ as small compared to the cut-off
$L$ . As a consequence it is easy to check that the contribution
of the first sum will indeed go to zero in the limit since
$\Lambda <<L$. In the large $L$ limit one can therefore conclude
that the full sum will be dominated by the second term
corresponding to high internal momenta $l_1$ and $l_2$, viz

\begin{eqnarray}
{\Delta}(p_1,p_2)&=&\sum_{\epsilon=-\frac{p_1-1}{2}}^{\frac{p_1-1}{2}}\sum_{l=\Lambda-\epsilon}^{L-\epsilon}\frac{(2l+1)^2-4{\epsilon}^2}{(l^2-{\epsilon}^2)((l+1)^2-{\epsilon}^2)}
    (L+1)\nonumber\\
    &{\times}&\left\{\begin{array}{ccc}
        l & \frac{L}{2}+\epsilon &\frac{L}{2}\\
     \frac{L}{2}-\epsilon & l &p_1\end{array}\right\} \left\{\begin{array}{ccc}
         l & \frac{L}{2}-\epsilon & \frac{L}{2}\\
     \frac{L}{2}+\epsilon & l
    & p_2
    \end{array}\right\}X(l_1,l_2,p_1,p_2),\nonumber\\
\end{eqnarray}
where for convenience we have also made  the following change of
variables $l=\frac{1}{2}(l_1+l_2)$,
$\epsilon=\frac{1}{2}(l_1-l_2){\equiv}\frac{n_1^{'}}{2}$ and then used Regge symmetries of
the $6j$ symbols . Since $\frac{L}{2}$, $\frac{L}{2}{\pm}\epsilon $
and $l>>p_1,-2{\epsilon}$ we can use for the asymptotic behaviour
of the $6j$ symbols the Edmonds' Formula \cite{VKM}, namely
\begin{eqnarray}
\left\{\begin{array}{ccc}
        l & \frac{L}{2}+\epsilon &\frac{L}{2}\\
     \frac{L}{2}-\epsilon & l
     &p_1\end{array}\right\}=\frac{1}{2}\frac{(-1)^{L+p_1+l }}{\sqrt{2l+1}}\bigg[\frac{(-1)^{-\epsilon}}{\sqrt{(L+1+2\epsilon
     )}}d_{-n_1^{'},0}^{p_1}(\theta)+\frac{(-1)^{\epsilon}}{\sqrt{(L+1-2\epsilon
     )}}d_{n_1^{'},0}^{p_1}(\theta)\bigg].\nonumber\\
\end{eqnarray}
where $d_{-n_1^{'},0}^{p_1}(\theta)$, $d_{n_1^{'},0}^{p_1}(\theta)$ are rotation matrices and
$\theta$ is the angle defined by
\begin{eqnarray}
cos{\theta}=\frac{l(l+1)+L\epsilon +\epsilon(\epsilon
+1)}{\sqrt{l(l+1)(L+2\epsilon)(L+2\epsilon +1)}}{\simeq}\frac{l}{L}~,~{\rm for}~L{\longrightarrow}{\infty}.
\end{eqnarray}
In the continuum large $L$ limit it is not difficult therefore to conclude that the leading contribution to the above loop takes the form
\begin{eqnarray}
{\Delta }(p_1,p_2)&=&\sum_{l=\Lambda }^{L}\frac{1}{2l^3}\sum_{n_1^{'}=-p_1}^{p_1}\big((-1)^{n_1^{'}}-(-1)^{p_1}\big)\bigg[(-1)^{n_1^{'}}d_{n_1^{'},0}^{p_1}(\theta)d_{n_1^{'},0}^{p_2}(\theta)+d_{-n_1^{'},0}^{p_1}(\theta)d_{n_1^{'},0}^{p_2}(\theta)\bigg]\nonumber\\
&\times& X(l,n_1^{'},p_1,p_2).
\end{eqnarray}
Without any loss of generality we have also assumed for simplicity that $X(l,-n_1^{'},p_1,p_2)=X(l,n_1^{'},p_1,p_2)$. Now we use the result
\begin{eqnarray}
d_{n_1^{'},0}^{p_2}(\theta)=\sqrt{\frac{4{\pi}}{2p_2+1}}(-1)^{n_1^{'}}Y_{p_2-n_1^{'} }(\theta,0)~,
\end{eqnarray}
and the fact that for $\phi =0$  we have $Y_{p-n}=(-1)^nY_{pn}$, $Y_{pn}^{*}=Y_{pn}$ to rewrite the above result in the equivalent form
\begin{eqnarray}
{\Delta}(p_1,p_2)&=&\frac{4{\pi}}{\sqrt{(2p_1+1)(2p_2+1)}}\sum_{l=\Lambda }^{L}\frac{1}{l^3}\sum_{n_1^{'}=-p_1}^{p_1}\big(1-(-1)^{n_1^{'}+p_1}\big)Y_{p_1n_1^{'}}(\theta,0)Y_{p_2n_1^{'}}^{*}(\theta,0)\nonumber\\
&\times& X(l,n_1^{'},p_1,p_2)\nonumber\\
&=&\frac{1}{L^2}\frac{4{\pi}}{\sqrt{(2p_1+1)(2p_2+1)}}\int_0^{\frac{\pi}{2}-\delta}\frac{\sin\theta d{\theta}}{\cos^3{\theta}}\sum_{n_1^{'}=-p_1}^{p_1}\big(1-(-1)^{n_1^{'}+p_1}\big)Y_{p_1n_1^{'}}(\theta,0)Y^{*}_{p_2n_1^{'}}(\theta,0)\nonumber\\
&\times&X(l,n_1^{'},p_1,p_2).\label{5.53}
\end{eqnarray}
In above we have replaced the sum $\sum_l$ by an integral $\int
dl$ and then made the change of variable $l=L\cos\theta$. In
particular the value $l=\Lambda-\epsilon$  corresponds to the
angle $\theta$ such that $\cos\theta {\simeq}\frac{\Lambda}{L}$,
i.e $\theta=\frac{\pi}{2}-\delta$ where
${\delta}=\frac{\Lambda}{L}$ while the value $l=L-\epsilon $
corresponds to the angle $\theta=0$. At this stage we must also
use the explicit expressions of the  functions $X$. By
inspection all these functions are  found to be dominated in the
large $L$ limit by quadratic terms in the variable
$l=L\cos{\theta}$ for which the above loop is finite and non-zero
as one might easily check. Contribution of the linear and
constant terms in $X$ are found to be vanishingly small in the
limit. We skip the corresponding trivial algebra for simplicity.
This result means in particular that  the operator ${\Delta}_F$ approaches zero in the continuum limit whereas the action
${\Gamma}_2^{(3A_2)}$ survives the continuum limit and thus it
provides non-trivial highly non-local interactions  between the
scalar field and the $2-$dimensional gauge field.

\section{Scalar effective potential and phase transition on ${\bf S}^2_L$}
In this section we will show explicitly and in some detail that the limiting theory $L={\infty}$ is actually independent of the gauge fixing parameter ${\xi}$ (see equation (\ref{ghost})) and hence the results obtained in the previous sections are gauge-invariant. As a result of this analysis we will also be able to identify a novel non-perturbative phase transition which occurs in the model. The effective action for generic values of ${\xi}$ reads
\begin{eqnarray}
{\Gamma}[D_a]&=&S[D_a]+\frac{1}{2}Tr_{3}TR\log{\Omega}_{\xi}-TR\log{\cal
D}^2\label{effectivexi}
\end{eqnarray}
where now
\begin{eqnarray}
({\Omega}_{\xi})_{ab}&=&\alpha {\delta}_{ab}+ {\cal D
}_c^2{\delta}_{ab}-(1+\frac{M^2}{|L|^2}-\frac{1}{\xi}){\cal D}_a{\cal D}_b+2{\cal
F}_{ab}+\frac{2M^2}{|L|^2}(D_c^2-|L|^2){\delta}_{ab}+\frac{4M^2}{|L|^2}D_aD_b\nonumber\\
\end{eqnarray}
As before we will set for simplicity $\alpha =0$ and study the two limits $M{\longrightarrow}0$ (which corresponds to the ARS action) and $M{\longrightarrow}{\infty}$ (which corresponds to projecting out the normal component of the gauge field from the theory). To simplify further we will only be interested in computing the effective potential of the scalar component of the gauge field which is in fact motivated by two other reasons. First the fact that tadpole diagrams of the model (\ref{actionbeta})  are largely controlled by the fluctuation of this scalar field. The second motivation is the fact that the expectation value of this scalar field provides a measure of the radius of the sphere (as we will explain) and thus a zero expectation value means that the sphere has disappeared and a phase transition occured. We believe that this result reported first in \cite{jun} can be captured in one-loop perturbation theory since the one-loop result becomes exact in the large $L$ limit as we will also argue in the following.

In the continuum theory the normal scalar field is defined in
terms of arbitrary gauge configurations $A_a$ by the linear
equation ${\Phi}=A_an_a$. It is not difficult to check that for a
constant normal scalar field we have $A_a={\rho}n_a$, i.e
${\Phi}{\equiv}{\rho}$  where ${\rho}$ is some constant real
number, with a curvature given by $
F_{ab}=i{\rho}{\epsilon}_{abc}n_c$.
The YM and CS actions for this configuration are both equal to
$\frac{1}{2g^2}{\rho}^2$ while the scalar action $S_{M}$ is
given by $\frac{2M^2}{g^2}{\rho}^2$. The classical continuum effective potential is
therefore given by
\begin{eqnarray}
U=\frac{1+2M^2}{g^2}{\rho}^2.
\end{eqnarray}
On the other hand and as we have explained earlier the normal
scalar field on the fuzzy sphere is defined in terms of arbitrary
gauge configurations $A_a$ by the quadratic equation
${\Phi}=\frac{1}{2}(x_aA_a+A_ax_a+\frac{A_a^2}{|L|})$. However a
constant scalar field configuration on the fuzzy sphere can still
be defined by $ A_a={\rho}x_a$ where ${\rho}$ is again  some
constant real number. This is because the corresponding curvature
is given by a similar equation with a correct
continuum limit, viz $
F_{ab}=i{\rho}(1+\frac{{\rho}}{|L|}){\epsilon}_{abc}x_c$.
In this case the normal scalar field ${\Phi}$ must also satisfy $
{\Phi}={\rho}+\frac{{\rho}^2}{2|L|}$.
This can be solved for $\rho$ in terms of $\Phi$ and one finds two
solutions, namely $
\frac{{\rho}_{\pm}}{|L|}=-1{\pm}\sqrt{1+\frac{2\Phi}{|L|}}$.
Clearly in the large $L$ limit one of the solutions converge to
the actual scalar field $\Phi$ whereas the other one diverges,
namely ${\rho}_{+}{\longrightarrow}\Phi$,
${\rho}_{-}{\longrightarrow}-2|L|$, i.e on the fuzzy sphere it seems
that for every value of $\Phi$ we can have two
different values of ${\rho}$. Obviously around the second solution ${\rho}_{-}$ perturbation theory can not be trusted. The sum $U$ of the Yang-Mills, Chern-Simons and
scalar actions for this configuration is given by  (with $\phi =1+\frac{\rho}{|L|}$)
\begin{eqnarray}
V=\frac{U}{|L|^2}=\frac{1}{2g^2}\left[\phi^4-\frac{4}{3}\phi^3+M^2\left(\phi^2-1\right)^2+\alpha\phi^2\right].
\end{eqnarray}
Quantum corrections can also be computed using equation (\ref{effectivexi}) for generic values of $\alpha$, $M$ and $\xi$ and one finds
the one-loop complete result
\begin{eqnarray}
U_{eff}&=&U+
\frac{1}{2}Tr_3 TR\log\bigg[\alpha\delta_{ab}+\phi^2 \big({\cal L}^2\delta_{ab}-\big[1+\frac{M^2}{|L|^2}-\frac{1}{\xi}\big]{\cal L}_{a}{\cal L}_{b}\big)+2\phi (1-\phi)(\vec{\theta}.\vec{\cal L})_{ab}\nonumber\\
& +& 2M^2(\phi^2-1)\delta_{ab}+4M^2\phi^2P^{N}_{ab}\bigg]-TR\log[{\phi}^2{\cal L}^2].
\end{eqnarray}
In above we have used the following identity ${\cal L}_a{\cal L}_b={\cal L}^2{\delta}_{ab}-(\vec{\theta}.\vec{\cal L})_{ab}-(\vec{\theta}.\vec{\cal L})^2_{ab}$ where ${\theta}_a$ are the generators of $SU(2)$ in the spin $1$ irreducible representation , i.e $[{\theta}_a,{\theta}_b]=i{\epsilon}_{abc}{\theta}_c$ , $\sum_a {\theta}_a^2=2$ , $({\theta}_a)_{bc}=-i{\epsilon}_{abc}$ . The total angular momentum is therefore $\vec{J}=\vec{\cal L}+\vec{\theta}$ .

For $M=0$ and $\alpha =0$ the classical potential takes the simple form

\begin{equation}
V[\phi]{\equiv}\frac{U[\phi]}{|L|^2}=\frac{1}{2g^2}{\phi}^{3}\left(\phi-\frac{4}{3}\right).
\end{equation}
The minimum of this potential is clearly $\phi=1$, i.e the fuzzy sphere with coordinates $D_a=L_a$. Quantum corrections are given by
\begin{eqnarray}
V_{eff}[\phi]=V[\phi]+4\log{\phi}+{\Delta}V
\end{eqnarray}
where
\begin{eqnarray}
{\Delta}V&=&\frac{1}{2|L|^2}Tr_3TR\log{\cal H}
-\frac{1}{|L|^2}TR\log{\cal L}^2\nonumber\\
{\cal H}&=&\bigg[{\cal L}^2+(\frac{1}{\phi}-1)(J^2-{\cal L}^2-2)+(1-\frac{1}{\xi})\big[\frac{1}{4}(J^2-{\cal L}^2)^2-\frac{1}{2}(J^2-{\cal L}^2)-{\cal L}^2\big]\bigg].
\end{eqnarray}
The eigenvalues of ${\cal L}^2$ and $\vec{J}^2$ are given respectively 
by $l(l+1)$ and $j(j+1)$ respectively where $l=1,...,L$ and $j=l+1,l,l-1$ . The corresponding eigenvectors are the vector spherical harmonics operators . Clearly the operator ${\cal H}$ is diagonal in this basis , explicitly we have
{\small{\begin{eqnarray}
{\Delta}V&=&\frac{1}{2|L|^2}\sum_l (2l+1) \log\big[\frac{1}{\xi}l(l+1)-2(\frac{1}{\phi}-1)\big]+\frac{1}{2|L|^2}\sum_{l}(2l+3)\log\big[l(l+1)+2l(\frac{1}{\phi}-1)\big]\nonumber\\
&+&\frac{1}{2|L|^2}\sum_{l}(2l-1)\log\big[l(l+1)-2(l+1)(\frac{1}{\phi}-1)\big]-\sum_{l}(2l+1)\log l(l+1).\nonumber
\end{eqnarray}}}
Obviously all ${\xi}-$dependence of this potential ${\Delta}V$ is confined to the first term. In the large $L$ limit the relevant terms are given by
\begin{eqnarray}
{\Delta}V&=&\frac{1}{2|L|^2}\sum_l (2l+1) \log\big[1-\frac{2\xi}{l(l+1)}(\frac{1}{\phi}-1)\big]+\frac{1}{2|L|^2}\sum_{l}(2l-1)\log\big[1-\frac{2}{l}(\frac{1}{\phi}-1)\big]\nonumber\\
&+&\frac{1}{2|L|^2}\sum_{l}(2l+3)\log\big[1+\frac{2}{l+1}(\frac{1}{\phi}-1)\big].\label{ll}\end{eqnarray}
It is rather a straightforward exercise to show that in the strict limit this potential ${\Delta}V$ vanishes as $\frac{\log L}{L^2}$. In other words the ${\xi}-$dependence drops completely and the theory $L={\infty}$ is gauge invariant. We end up therefore with the simple potential 
\begin{eqnarray}
V_{eff}=\frac{1}{2g^2}({\phi}^4-\frac{4}{3}{\phi}^3)+4\log \phi.
\end{eqnarray}
Taking the first and second derivatives of this potential we obtain
\begin{eqnarray}
&&V_{eff}^{'}=\frac{1}{2g^2}(4{\phi}^3-4{\phi}^2)+\frac{4}{\phi}\nonumber\\
&&V_{eff}^{''}=\frac{1}{2g^2}(12{\phi}^2-8{\phi})-\frac{4}{{\phi}^2}.
\end{eqnarray}
The condition $V^{'}(\phi)=0$ which also read
\begin{eqnarray}
{\phi}^4-{\phi}^3+2g^2=0
\end{eqnarray}
 will give us extrema of the model. These extrema are minima and thus stable if the condition $V^{''}(\phi)>0$ (or equivalently $3{\phi}^4-2{\phi}^3-2g^2>0$) is satisfied whereas they are maxima if $3{\phi}^4-2{\phi}^3-2g^2<0$. The equation
\begin{eqnarray}
3{\phi}^4-2{\phi}^3-2g^2=0
\end{eqnarray}
tell us therefore exactly when we go from a bounded potential to an unbounded potential (see the attached figure). Solving the above two equations yield immediately the minimum ${\phi}{\equiv}{\phi}_{\infty}=\frac{3}{4}$ with the corresponding critical value
\begin{eqnarray}
g^2{\equiv}g^2_{*}=\frac{1}{8}\left(\frac{3}{4}\right)^3.\label{cv}
\end{eqnarray}
\begin{figure}[h]
\begin{center}
\includegraphics[width=9cm]{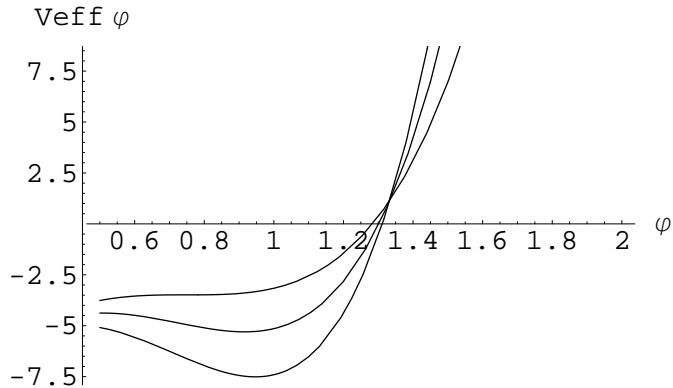}
\caption{ Effective potential for different values of g}
\end{center}
\end{figure}
This agrees nicely with the result of \cite{jun} which was however obtained by simulating the full theory. In other words the one-loop calculation of this article reproduces their exact result. This is anyway expected because of the following simple reason. It is easily seen that in this $U(1)$  model all vertices are given in terms of commutators and thus they all vanish in the continuum limit . Hence since  we know that in  $2-$dimensional gauge models only one-loop diagrams can diverge we  can conclude that higher loops must strictly vanish in the continuum limit and thus they do not contribute. 

We can  clearly see from the above result that at the critical value (\ref{cv}) a first order phase transition occur which separates the fuzzy sphere phase from the pure matrix phase. The fuzzy sphere phase is where the interpretation that we have a $U(1)$ gauge theory on a ( fuzzy or continuum ) sphere is valid and it holds for $g<g_{*}$. The matrix phase is where this interpretation brakes down and it holds for $g>g_{*}$. Beyond this critical point the fuzzy sphere seems to disappear as the radius goes to zero (the radius here is identified with the inverse of the order parameter $\phi$ which diverges for $g>g_{*}$). We should also note here that the one-loop approximation imposes a minimum value for $\phi$ given by $\phi=\frac{2}{3}$. The leading contributions in the large $L$ limit of this one-loop result can also be computed (say) from (\ref{ll}) but we postpone this as well as a thorough discussion of the phase diagram to a future communication \cite{future} where we will also provide an alternative derivation of this phase transition.

\section{The $U(1)$ theory with a large mass term.}
\subsection{The quadratic effective action in the limit $M{\longrightarrow}{\infty}$.} In the presence of a mass
term for the scalar field the quadratic action (\ref{lll})
becomes
\begin{eqnarray}
S_2=-\frac{1}{2g^2}Tr_L[L_a,A_b]^2+\frac{1}{2g^2}Tr_L[L_a,A_a]^2+\frac{M^2}{2g^2}Tr_L(x_aA_a+A_ax_a)^2\label{lllM}
\end{eqnarray}
Whereas the quadratic effective action (\ref{ex}) becomes now
given by
{\small{\begin{eqnarray}
{\Gamma}_2  &=&S_2+TR\bigg(-\frac{1}{2}
\bigg(\log\frac{1}{{\Delta}}\bigg)_{aa}+\log\frac{1}{{\cal
L}^2}\bigg)+TR\bigg(\frac{1}{2}\bigg(\frac{1}{\Delta}\bigg)_{ab}{\omega}^{(1)}_{ba}-\frac{1}{{\cal
L}^2}({\cal L}{\cal A}+{\cal A}{\cal L})\bigg)\nonumber\\
&+&TR\bigg(\frac{1}{2}\bigg(\frac{1}{\Delta}\bigg)_{ab}{\omega}_{ba}^{(2)}-\frac{1}{4}\bigg(\frac{1}{\Delta}\bigg)_{ac}{\omega}_{cd}^{(1)}\bigg(\frac{1}{\Delta}\bigg)_{db}{\omega}^{(1)}_{ba}
-\frac{1}{{\cal L}^2}{\cal A}^2+\frac{1}{2}\frac{1}{{\cal
L}^2}({\cal L}{\cal A}+{\cal A}{\cal L})\frac{1}{{\cal
L}^2}({\cal L}{\cal A}+{\cal A}{\cal L})\bigg) .\label{exM}\nonumber\\
\end{eqnarray}}}
The propagator of the gauge degrees of freedom is given here in
terms of the Laplacian ${\Delta}={\cal L}^2+4M^2P^N$ where $P^N$
is the normal projector (\ref{normal}) and as a consequence the
calculation of the above effective action is now more complicated
and involves the propagator
\begin{eqnarray}
\frac{1}{\Delta}=\frac{1}{{\Delta}_0}+\frac{1}{M^2}\frac{1}{{\delta}{\Delta}}~,~\frac{1}{{\delta}{\Delta}}=M^2\big(-\frac{1}{{\Delta}_0}V\frac{1}{{\Delta}_0}+\frac{1}{{\Delta}_0}V\frac{1}{{\Delta}_0}V\frac{1}{{\Delta}_0}+...\big).
\end{eqnarray}
where $\frac{1}{{\Delta}_0}$ is the inverse of the diagonal
Laplacian 
\begin{eqnarray}
{\Delta}_0=P^T{\cal L}^2P^T+P^N({\cal L}^2+4M^2)P^N
\end{eqnarray}
and hence $\frac{1}{{\Delta}_0}=P^T\frac{1}{{\cal
L}^2}P^T+P^N\frac{1}{{\cal L}^2+4M^2}P^N$  whereas the vertex $V$
is the off diagonal part of the Laplacian ${\Delta}$, i.e
\begin{eqnarray}
V=P^T{\cal L}^2P^N+P^N{\cal L}^2P^T. 
\end{eqnarray}
In (\ref{exM}) the first
quantum correction is just a number which is irrelevant in any
case, the second correction gives the tadpole diagrams of Figure
$1$ whereas the third correction gives the vacuum polarization
diagrams of Figures $2$ and $3$ where the operators
${\omega}^{(1)}$ and ${\omega}^{(2)}$ are given explicitly by

\begin{eqnarray}
{\omega}^{(1)}_{ab}&=&({\cal L}{\cal A}+{\cal A}{\cal
L}){\delta}_{ab}+2{\cal
F}^{(0)}_{ab}+\frac{2M^2}{|L|^2}(LA+AL){\delta}_{ab}+\frac{4M^2}{|L|^2}(L_aA_b+A_aL_b)\nonumber\\
{\omega}^{(2)}_{ab}&=&{\cal A}^2{\delta}_{ab}+2[{\cal A}_a,{\cal
A}_b]+\frac{2M^2}{|L|^2}A^2{\delta}_{ab}+\frac{4M^2}{|L|^2}A_aA_b.
\end{eqnarray}
Our interest is to compute this effective action (\ref{exM}) in
the limit where we take $M{\longrightarrow}{\infty}$ first then
$L{\longrightarrow}{\infty}$. To this end we combine the tadpole
diagrams ( Figure $1$ ) and the diagrams corresponding to the
$4-$vertex correction to the vacuum polarization tensor (Figure
$2$) and write them in the form
\begin{eqnarray}
{\Gamma}^M_1+{\Gamma}^{(4)M}_2&=&\frac{1}{2}TR\bigg[\big(\frac{1}{\Delta}\big)_{ab}\bigg(({\cal
L}{\cal A}+{\cal A}{\cal L}+{\cal A}^2\big){\delta}_{ab}-2{\cal
F}_{ab}+4\frac{M^2}{|L|^2}(L_bA_a+A_bL_a+A_bA_a)\nonumber\\
&+&4\frac{M^2}{|L|}\phi{\delta}_{ab}\bigg)\bigg]
-TR\frac{1}{{\cal L}^2}\big({\cal L}{\cal A}+{\cal A}{\cal
L}+{\cal A}^2 \big).\label{148}
\end{eqnarray}
There are in total $8$ terms to be computed in the large mass
limit before we take the actual continuum limit. As it turns out
it is enough to compute these terms only and then use the
requirement of gauge invariance to infer the structure of the
$3-$vertex correction to the vacuum polarization tensor (Figure
$3$). These are obviously given now by
\begin{eqnarray}
{\Gamma}^{(3F)M}_2+{\Gamma}^{(3A)M}_2=TR\bigg(-\frac{1}{4}\bigg(\frac{1}{\Delta}\bigg)_{ac}{\omega}_{cd}^{(1)}\bigg(\frac{1}{\Delta}\bigg)_{db}{\omega}^{(1)}_{ba}
+\frac{1}{2}\frac{1}{{\cal L}^2}({\cal L}{\cal A}+{\cal A}{\cal
L})\frac{1}{{\cal L}^2}({\cal L}{\cal A}+{\cal A}{\cal L})\bigg) .
\end{eqnarray}
So in the following we will only compute the action (\ref{148})
explicitly. In the large $M$ limit we will also make use of the
fact that we have
\begin{eqnarray}
&&\frac{1}{{\Delta}_0}{\longrightarrow}P^T\frac{1}{{\cal
L}^2}P^T+O(\frac{1}{M^2}),
\end{eqnarray}
and
\begin{eqnarray}
\frac{1}{{\delta}{\Delta}}{\longrightarrow}-\frac{1}{4}P^T\frac{1}{{\cal
L}^2}P^T{\cal L}^2P^N-\frac{1}{4}P^N{\cal L}^2P^T\frac{1}{{\cal
L}^2}P^T+\frac{1}{4}P^T\frac{1}{{\cal L}^2}P^T{\cal L}^2P^N{\cal
L}^2P^T\frac{1}{{\cal L}^2}P^T+O(\frac{1}{M^2}).\label{lpk}\nonumber\\
\end{eqnarray}
The contribution (\ref{148}) decomposes into the sum of two parts, namely
\begin{eqnarray}
&&\frac{1}{2}TR\bigg[\big(\frac{1}{\Delta}_0\big)_{ab}\bigg(({\cal
L}{\cal A}+{\cal A}{\cal L}+{\cal A}^2\big){\delta}_{ab}-2{\cal
F}_{ab}+4\frac{M^2}{|L|^2}(L_bA_a+A_bL_a+A_bA_a)+4\frac{M^2}{|L|}\phi{\delta}_{ab}\bigg)\bigg]\nonumber\\
&&-TR\frac{1}{{\cal
L}^2}\big({\cal L}{\cal A}+{\cal A}{\cal L}+{\cal A}^2
\big)\label{sss}
\end{eqnarray}
and
\begin{eqnarray}
\frac{2}{|L|^2}TR\bigg(\frac{1}{{\delta}{\Delta}}\bigg)_{ab}\big[L_bA_a+A_bL_a+A_bA_a+|L|\phi
{\delta}_{ab}\big]{\equiv}\frac{2}{|L|^2}TR
\bigg(\frac{1}{{\delta}{\Delta}}\bigg)_{ab}\big[D_bD_a+|L|\phi
{\delta}_{ab}\big],\label{79}
\end{eqnarray}
where we have also used in the above last equation (\ref{79}) the
fact that $ TR(\frac{1}{{\delta}{\Delta}})_{ab}L_bL_a=0$ which
can be easily seen from the expression (\ref{lpk}). A long
straightforward calculation shows that in the large $L$ limit the
contribution (\ref{sss}) depends only on the scalar component of
the gauge field, i.e (\ref{sss}) becomes a simple scalar action
given explicitly by
\begin{eqnarray}
3L\int \frac{d{\Omega}}{4{\pi}}\phi +2 \int
\frac{d{\Omega}}{4{\pi}}{\phi}^2.
\end{eqnarray}
By construction the second term of (\ref{79}) can only depend on
the scalar field and thus if it does not vanish in the continuum
limit it will at most provide an extra scalar quantum
contribution. Hence we do not really need to compute it explicitly. However as a final exercise we compute in some detail  the first
term of (\ref{79}). By using te equation (\ref{identity1}) and (\ref{identity2}) immediately
we can compute that
\begin{eqnarray}
\frac{2}{|L|^2}TR
\bigg(\frac{1}{{\delta}{\Delta}}\bigg)_{ab}D_bD_a&=&-\frac{1}{2|L|^2}\sum_{lm}\frac{1}{l(l+1)}Tr_L\hat{Y}_{lm}{\cal
L}^2(\hat{Y}_{lm}^{+}P^T_{ab})P^N_{bc}D_cA_dP_{da}^T\nonumber\\
&-&\frac{1}{2|L|^2}\sum_{lm}\frac{1}{l(l+1)}Tr_L{\cal
L}^2(P^T_{ab}\hat{Y}_{lm}^{+})\hat{Y}_{lm}P^T_{bc}A_cD_dP_{da}^N\nonumber\\
&+&\frac{1}{2|L|^2}\sum_{lm,kn}\frac{1}{l(l+1)}\frac{1}{k(k+1)}\big(Tr_L\hat{Y}_{lm}\hat{Y}_{kn}^{+}P^T_{fb}A_bA_aP_{ac}^T\big)\nonumber\\
&{\times}&\big(Tr_L{\cal
L}^2(\hat{Y}_{lm}^{+}P^T_{cd})P^N_{de}{\cal
L}^2(P^T_{ef}\hat{Y}_{kn})\big).\label{80}\nonumber\\
\end{eqnarray}
It is rather trivial to show that the first two terms above
vanish in the continuum large $L$ limit and as  a consequence we
end up effectively with the expression
\begin{eqnarray}
\frac{2}{|L|^2}TR
\bigg(\frac{1}{{\delta}{\Delta}}\bigg)_{ab}D_bD_a=\frac{1}{2|L|^2}Tr_L
(A_c^T)^{+}M_{cf}A_f^T~,\label{81}
\end{eqnarray}
where
\begin{eqnarray}
 ~M_{cf}=\sum_{lm,kn}\frac{1}{l(l+1)}\frac{1}{k(k+1)}\bigg(Tr_L{\cal
L}^2(\hat{Y}_{lm}^{+}P^T_{cd})P^N_{de}{\cal
L}^2(P^T_{ef}\hat{Y}_{kn})\bigg)\hat{Y}_{lm}\hat{Y}_{kn}^{+},
\end{eqnarray}
and where we have also defined the tangent gauge field
$A_a^T=P_{ab}^TA_b$. By construction the matrix $M$ can only be
proportional to the identity matrix, i.e
$M_{cf}=\frac{1}{3}M_{aa}{\delta}_{cf}$, and furthermore we can
show that in the continuum limit the trace $M_{aa}$    vanishes
identically as $\frac{1}{L}$. This can be easily derived by using
the fact that ${\cal L}^2(\hat{Y}_{lm}^{+}P^T_{cd})P^N_{de}{\cal
L}^2(P^T_{ef}\hat{Y}_{kn}){\longrightarrow}16\hat{Y}_{lm}^{+}\hat{Y}_{kn}x_cx_f+8\hat{Y}_{lm}^{+}{\cal
L}_A(\hat{Y}_{kn}){\cal L}_A(x_cx_f)+4{\cal
L}_A(\hat{Y}_{lm}^{+}){\cal L}_B(\hat{Y}_{kn}){\cal
L}_A(x_c){\cal L}_B(x_f)$ when $L{\longrightarrow}{\infty}$. Hence
the contribution (\ref{81}) vanishes  in this continuum limit.

Putting all these results together we conclude that the action
(\ref{148}), which gives the sum of the tadpole diagrams (Figure
$1$) and the $4-$vertex correction to the vacuum polarization
tensor (Figure $2$), becomes a purely scalar action in the
limit where we take $M{\longrightarrow}{\infty}$ first and then
$L{\longrightarrow}{\infty}$, i.e
\begin{eqnarray}
{\Gamma}^M_1+{\Gamma}^{(4)M}_2&=&3L\int
\frac{d{\Omega}}{4{\pi}}\phi +2 \int
\frac{d{\Omega}}{4{\pi}}{\phi}^2+\frac{2}{|L|}TR
\bigg(\frac{1}{{\delta}{\Delta}}\bigg)_{aa}\phi.\label{rodrigo}
\end{eqnarray}
 This pure scalar action should  be compared with the
continuum limit of the sum of the two actions (\ref{5.21}) and
(\ref{5.26}) which as we have shown does depend in the continuum
limit on the $2-$dimensional gauge field. In other words
suppressing the normal component of the gauge field by giving it
a large mass allowed us to suppress in the limit the contribution
of the tangential field to the tadpole and to the $4-$vertex
correction of the vacuum polarization tensor. By the requirement
of gauge-invariance this suppression will also occur in the other
contributions to the vacuum polarization tensor and as
consequence the large mass of the scalar field regulates
effectively the UV-IR mixing which is consistent with
\cite{stein1}.

\subsection{The phase transition in the limit $M{\longrightarrow}{\infty}$}
We note here that the above phase transition can also be thought of as a non-perturbative representation of the UV-IR mixing phenomena. In order to see this more clearly we must take the effect of the mass into account and then calculate how the critical coupling constant scales when we start increasing $M$. Since the UV-IR mixing  disappears in the large mass limit we must be able to see that the phase transition becomes harder to reach from small couplings when we increase $M$. This is indeed the case as we will now show.
In the case $M{\neq}0$, the classical and the effective actions are given respectively by

\begin{eqnarray}
V=\frac{U}{|L|^2}=\frac{1}{2g^2}\left[\phi^4-\frac{4}{3}\phi^3+M^2\left(\phi^2-1\right)^2\right]
\end{eqnarray}
and
\begin{eqnarray}
V_{Meff}[\phi]=V_M[\phi]+4\log\phi+{\Delta}V_M
\end{eqnarray}
where
\begin{eqnarray}
{\Delta}V_M&=&\frac{1}{2|L|^2}Tr_3TR\log({\cal H}_M+4M^2P^N)
-\frac{1}{|L|^2}TR\log\mathcal{L}^2
\end{eqnarray}
and
\begin{eqnarray}
{\cal H}_M&=&{\cal L}^2+(\frac{1}{\phi}-1)(J^2-{\cal L}^2-2)+(1+\frac{M^2}{|L|^2}-\frac{1}{\xi})\big[\frac{1}{4}(J^2-{\cal L}^2)^2-\frac{1}{2}(J^2-{\cal L}^2)-{\cal L}^2\big]\nonumber\\
&+&2M^2(1-\frac{1}{{\phi}^2}).
\end{eqnarray}
As before we can argue that in the large $L$ limit the relevant terms in the effective potential are given  by

\begin{eqnarray}
V_{Meff}[\phi]=V_M[\phi]+4\log\phi.
\end{eqnarray}
Solving for the critical values using the same method outlined in a previous subsection yields the results
\begin{eqnarray}
{\phi}_{*}^{\pm}=\frac{3}{8(1+M^2)}\bigg[1{\pm}\sqrt{1+\frac{32M^2(1+M^2)}{9}}\bigg].
\end{eqnarray}
And
\begin{eqnarray}
g_{*}^2=-\frac{1}{2}(1+M^2){\phi}_{*}^4+\frac{1}{2}{\phi}_{*}^3+\frac{M^2}{2}{\phi}_{*}^2.
\end{eqnarray}
Extrapolating to large masses  ($M{\longrightarrow}{\infty}$) we obtain the scaling behaviour 
\begin{eqnarray}
&&{\phi}_{*}^{\pm}{\longrightarrow}{\pm}\frac{1}{\sqrt{2}}\nonumber\\
&&g_{*}^2{\longrightarrow}\frac{M^2}{8}.
\end{eqnarray}

\begin{figure}[h]
\begin{center}
 \includegraphics[width=9cm]{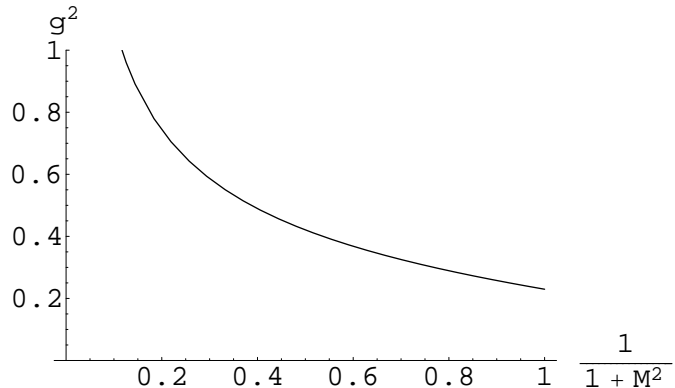} 
\caption{{ The phase diagram.}}
\end{center}
\end{figure}

In other words the phase transition happens each time at a larger value of the coupling constant when $M$ is increased and hence it is harder for the system to reach the pure matrix phase for large enough masses if one starts of course from the fuzzy sphere phase. Putting it differently we have virtually the fuzzy sphere interpretation at all scales of the coupling constant which is indeed what we want if this model is going to approximate a $U(1)$ gauge theory in two dimensions.

\newpage
\section{Conclusion}
The fuzzy sphere ${\bf S}^2_L$ is a noncommutative space with finite number of degrees of freedom. As we have discussed it is an approximation of the ordinary sphere ${\bf S}^2$ which leaves all commutative symmetries intact. Thus it is only natural to consider this space (instead of a naive lattice) as a nonperturbative regularization of chiral gauge theories in dimension two. Generalization to $4-$dimensions is straightforward where we use ${\bf S}^2_{L_1}{\times}{\bf S}^2_{L_2}$ as the underlying fuzzy regularization of Euclidean compact spacetime. 

It is always a good strategy to study the properties of perturbation theory on these fuzzy spaces in some detail before any serious simulation can be attempted. In particular there is this perturbative UV-IR mixing phenomena which is expected to manifest itself in some form or shape in all noncommutative (fuzzy or otherwise) field theories. The nonperturbative origin of this mixing is still unknown and a deeper interpretation of it is still lacking. Let us also recall here that the fuzzy spheres ${\bf S}^2_L$ and ${\bf S}^2_{L_1}{\times}{\bf S}^2_{L_2}$ can  be thought of as regularizations of the Moyal-Weyl planes ${\bf R}^2_{\theta}$ and ${\bf R}^2_{{\theta}_1}{\times}{\bf R}^2_{{\theta}_2}$ respectively. Thus understanding the UV-IR mixing in the (more conceptually simple) finite setting of the fuzzy spheres may give us a better understanding of the UV-IR mixing on the noncommutative planes.

 In this article we have considered for simplicity the case of a $U(1)$ gauge field on one single fuzzy sphere. Extension of this analysis to higher gauge groups $U(n)$ and to $4-$dimensions will be reported elsewhere
\cite{future}.

From a string theory point of view we know that the most natural gauge action on the fuzzy sphere is the  Alekseev-Recknagel-Schomerus action which is a particular combination of the Yang-Mills action and the Chern-Simons term. We computed the one-loop quadratic effective action and showed explicitly the existence of a gauge-invariant UV-IR mixing in the continuum limit $L{\longrightarrow}{\infty}$. In other words the quantum $U(1)$ effective action does not vanish in the commutative limit and a noncommutative anomaly survives. We computed also the scalar effective potential and proved the gauge-fixing-independence of the limiting model $L={\infty}$ and then showed explicitly that the one-loop result predicts  a first order phase transition which was observed recently in the simulation of \cite{jun}. The one-loop result for the $U(1)$ theory is therefore exact in this limit. 

Since the differential calculus on the fuzzy sphere is $3-$dimensional the model contains an extra scalar fluctuation which is normal to the sphere. We have  argued that if we add a large mass term for this scalar mode the UV-IR mixing will be completely removed from the gauge  sector. This is in accordance with the large $L$ analysis of the model done in \cite{stein1} and shows that the UV-IR mixing of this theory is only confined to the scalar sector. This argument falls however a little short of being a rigorous proof. Correspondingly the phase transition becomes harder to reach starting from small couplings when $M$ is increased. This suggests that the phase transition we are observing is a nonperturbative manifestation of the UV-IR mixing phenomena.

\paragraph{Acknowledgements}
The authors P. Castro-Villarreal and R. Delgadillo-Blando  would like to thank Denjoe O'Connor for his supervision through the course of this study. B. Ydri would like to thank  Denjoe O'Connor for his extensive discussions and critical comments while this research was in progress. B. Ydri would also like to thank A. P. Balachandran for his comments and suggestions. The work of P. C. V. and R. D. B. is supported by CONACyT M\'exico.

\appendix

\section{ Tadpole diagrams}
Quantum correction to the tree-level linear term $S_1=0$ is given
by the sum of the two tadpole diagrams of Figure $1$ , viz
\begin{eqnarray}
{\Gamma}_1 & = &
\frac{1}{2}TR\Delta^{(1)}=\frac{1}{2}TR\frac{1}{{\cal L}^2}({\cal
L}{\cal A}+{\cal A}{\cal L})\nonumber\\
&=&\frac{1}{2}\left(\frac{1}{{\cal
L}^2}\right)^{AB,CD}\bigg[({\cal L})^{CD,EF}({\cal
A})^{EF,AB}+({\cal A})^{CD,EF}({\cal L})^{EF,AB}\bigg].
\end{eqnarray}
The operators ${\cal L}_a(..)=[L_a,..]$ and ${\cal
A}_a(..)=[A_a,..]$ carry $4$ indices because they can act on
matrices either from the left or from the right . We use now the
identity
\begin{eqnarray}
\frac{1}{L+1}\sum_{lm}\hat{Y}_{lm}^{AB}(\hat{Y}_{lm}^{+})^{CD}=\delta^{AD}\delta^{BC},
\end{eqnarray}
to find the $2-$point Green's function
\begin{eqnarray}
\left(\frac{1}{\mathcal{L}^2}\right)^{AB,CD}=\frac{1}{L+1}\sum_{lm}\frac{\hat{Y}_{lm}^{AB}(\hat{Y}_{lm}^{+})^{DC}}{l(l+1)}.
\end{eqnarray}
Internal momenta will always be denoted by $(lm)$ while external momenta will be denoted by $(pn)$ . Thus
\begin{eqnarray}
\Gamma_1 & = &
-\sum_{lm}\frac{Tr_L[L_a,\hat{Y}_{lm}^{\dagger}][A_a,\hat{Y}_{lm}]}{l(l+1)}\nonumber\\
&=&-\sum_{pn}A_{-\mu}(pn)(-1)^{\mu}\sum_{lm}\frac{Tr_L[L_{\mu},\hat{Y}_{l-m}][\hat{Y}_{pn},\hat{Y}_{lm}]}{l(l+1)}.
\end{eqnarray}
In above we have used the fact that
$A_a={\eta}_a^{\mu}A_{\mu}$ where the coefficients
${\eta}_a^{\mu}$ satisfy
${\eta}_a^{\mu}{\eta}_a^{\nu}=(-1)^{\mu}{\delta}_{\mu+\nu,0}$ ,
$a=1,2,3$ , $\mu=0,+1,-1$. We use now the identities
\begin{eqnarray}
[L_{\mu},\hat{Y}_{lm}]=\sqrt{l(l+1)}C_{lm1\mu}^{lm+\mu}\hat{Y}_{lm+\mu}
\end{eqnarray}
\begin{eqnarray}
\hat{Y}_{l_1m_1}\hat{Y}_{l_2m_2}=\sqrt{L+1}\sqrt{(2l_1+1)(2l_2+1)}\sum_{l_3m_3}(-1)^{L+l_3}\left\{\begin{array}{ccc}
        l_1 & l_2 & l_3 \\
    \frac{L}{2} & \frac{L}{2} &\frac{L}{2}
    \end{array} \right\}C_{l_1m_1l_2m_2}^{l_3m_3}\hat{Y}_{l_3m_3},
\end{eqnarray}
and
\begin{eqnarray}
\sum_{m}C_{l-m1\mu}^{l-m+\mu}C_{pnlm}^{lm-\mu}=\frac{2l+1}{3}{\delta}_{p1}{\delta}_{n,-\mu}
\end{eqnarray}
\begin{eqnarray}
\left\{\begin{array}{ccc}
        1 & l & l \\
    \frac{L}{2} & \frac{L}{2} &\frac{L}{2}
    \end{array} \right\}=(-1)^{L+l+1}\frac{\sqrt{l(l+1)}}{\sqrt{2l+1}}\frac{1}{\sqrt{L+1}}\frac{1}{2|L|},
\end{eqnarray}
where  $C_{abcd}^{ef}$ are the standard Clebsch-Gordan coefficients and $\{\cdot\cdot\cdot\}$ are the standard $6j$ symbols \cite{VKM} to obtain the final result
\begin{eqnarray}
\sum_{m}(-1)^mTr_L[L_{\mu},\hat{Y}_{l-m}][\hat{Y}_{pn},\hat{Y}_{lm}]=-\frac{1}{\sqrt{3}}(-1)^{\mu}{\delta}_{p1}{\delta}_{n,-\mu}\frac{(2l+1)l(l+1)}{|L|}.
\end{eqnarray}
Tadpole diagrams are given therefore by
\begin{eqnarray}
{\Gamma}_1=\frac{4}{\sqrt{3}}|L|A_{-\mu}(1-\mu){\equiv}4|L|Tr_LA_ax_a.
\end{eqnarray}
By using the definition of the normal scalar field $\phi=\frac{1}{2}(x_aA_a+A_ax_a+\frac{A_a^2}{|L|})$ we rewrite the above action in the form (\ref{5.21}) .

\section{$4-$Vertex correction} This is given by
\begin{eqnarray}
{\Gamma}_2^{(4)} & = &
\frac{1}{2}{TR}\Delta^{(2)}=  -\frac{1}{2}\sum_{l_1m_1}\frac{Tr_L[A_a,\hat{Y}_{l_1m_1}^{\dagger}][A_a,\hat{Y}_{l_1m_1}]}{l_1(l_1+1)}\nonumber\\
& = &
-\frac{1}{2}\sum_{p_1n_1}\sum_{p_2n_2}A_{\mu}
(p_1n_1)A_{-\mu}(p_2n_2)(-1)^{\mu}\sum_{l_{1}m_{1}}(-1)^{m_1}\frac{Tr_L[\hat{Y}_{p_1n_1},\hat{Y}_{l_1-m_1}][\hat{Y}_{p_2n_2},\hat{Y}_{l_1m_1}] }{l_1(l_1+1)}.\nonumber\\
\end{eqnarray}
This corresponds to the combination of the two diagrams of Figure $5$ . The sum over $m_1$ can be done again by using now the identity
\begin{eqnarray}
\sum_{m_1m_2}(-1)^{m_1+m_2}C_{p_1n_1l_1-m_1}^{l_2m_2}C_{p_2n_2l_1m_1}^{l_2-m_2}=\frac{2l_2+1}{\sqrt{(2p_1+1)(2p_2+1)}}(-1)^{m_1}(-1)^{l_1+l_2+p_1}{\delta}_{p_1p_2}{\delta}_{n_1,-n_2}.\nonumber\\
\end{eqnarray}
Hence we obtain
\begin{eqnarray}
\sum_{m_1}(-1)^{m_1}Tr_L[\hat{Y}_{p_1n_1},\hat{Y}_{l_1-m_1}][\hat{Y}_{p_2n_2},\hat{Y}_{l_1m_1}]&=&-2(L+1)(2l_1+1){\delta}_{p_1p_2}{\delta}_{n_1,-n_2}(-1)^{n_1}
\sum_{l_2}(2l_2+1)\nonumber\\
&{\times}&(1-(-1)^{p_1+l_1+l_2})
\left\{\begin{array}{ccc}
        p_1 & l_1 & l_2 \\
    \frac{L}{2} & \frac{L}{2} & \frac{L}{2}
    \end{array} \right\}^2.
\end{eqnarray}
We get therefore the answer
\begin{eqnarray}
{\Delta}{\Gamma}_2^{(4)}=2\sum_{p_1n_1}|A_a(p_1n_1)|^2\sum_{l_1,l_2}\frac{2l_1+1}{l_1(l_1+1)}\frac{2l_2+1}{l_2(l_2+1)}(L+1)\left\{\begin{array}{ccc}
        p_1 & l_1 & l_2 \\
    \frac{L}{2} & \frac{L}{2} & \frac{L}{2}
    \end{array} \right\}^2l_2(l_2+1),
\end{eqnarray}
with the conservation law that $p_1+l_1+l_2$ is an odd number. In configuration space this contribution takes the form

\begin{eqnarray}
{\Delta}{\Gamma}_2^{(4)}=Tr_LA_a{\cal L}^2{\Delta}_4({\cal L}^2)A_a,
\end{eqnarray}
which is (\ref{5.26}) .

\section{The $3-$Vertex contribution}
In this appendix we calculate the Feynman diagrams $6a$ and $6b$ and. The contribution of the ${\cal F}$ term is given by the diagram of Figure $6b$, namely
\begin{eqnarray}
{\Gamma}_2^{(3F)}&=&-\frac{1}{4}Tr_3TR({\Delta}^{(j)})^2=TR\frac{1}{{\cal L}^2}{\cal F}_{ab}^{(0)}\frac{1}{{\cal L}^2}{\cal F}_{ab}^{(0)}\nonumber\\
&=&\sum_{k_1m_1}\sum_{k_2m_2}\frac{Tr_L\left[F_{ab}^{(0)}[\hat{Y}_{k_2m_2},\hat{Y}_{k_1m_1}^{+}]\right]Tr_L\left[F_{ab}^{(0)}[\hat{Y}_{k_1m_1},\hat{Y}^{+}_{k_2m_2}]\right]}{k_1(k_1+1)k_2(k_2+1)}\nonumber\\
&=&\sum_{p_1n_1}\sum_{p_2n_2}F_{ab}^{(0)}(p_1n_1)F_{ab}^{(0)}(p_2n_2)\sum_{k_1m_1k_2m_2}\frac{Tr_L\hat{Y}_{p_1n_1}[\hat{Y}_{k_2m_2},\hat{Y}_{k_1m_1}^{+}]Tr_L\hat{Y}_{p_2n_2}[\hat{Y}_{k_1m_1},\hat{Y}^{+}_{k_2m_2}]}{k_1(k_1+1)k_2(k_2+1)}\nonumber\\
\end{eqnarray}
The sum over $m_1$ and $m_2$ can be easily done and one obtains
\begin{eqnarray}
&&\sum_{m_1m_2}(-1)^{m_1+m_2}Tr_L\left[\hat{Y}_{p_1n_1}[\hat{Y}_{k_2m_2},\hat{Y}_{k_1-m_1}]\right]Tr_L\left[\hat{Y}_{p_2n_2}[\hat{Y}_{k_1m_1},\hat{Y}_{k_2-m_2}]\right]\nonumber\\
&=&2(2k_1+1)(2k_2+1){\delta}_{p_1p_2}{\delta}_{n_1,-n_2}(-1)^{n_1}
(L+1)\big(1
-(-1)^{k_1+k_2+p_1}\big)
\left\{\begin{array}{ccc}
        k_1 & k_2 & p_1 \\
    \frac{L}{2} & \frac{L}{2} & \frac{L}{2} \end{array}\right\}^2,\nonumber\\
\end{eqnarray}
and hence we get the contribution
\begin{eqnarray}
{\Gamma}_2^{(3F)}
&=&2\sum_{p_1n_1}|F_{ab}^{(0)}(p_1n_1)|^2\sum_{k_1k_2}\frac{2k_1+1}{k_1(k_1+1)}\frac{2k_2+1}{k_2(k_2+1)}(L+1)(1-(-1)^{k_1+k_2+p_1})\left\{\begin{array}{ccc}
        k_1 & k_2 & p_1 \\
    \frac{L}{2} & \frac{L}{2} & \frac{L}{2} \end{array}\right\}^2\nonumber\\
&=&4\sum_{p_1n_1}|F_{ab}^{(0)}(p_1n_1)|^2\sum_{k_1k_2}\frac{2k_1+1}{k_1(k_1+1)}\frac{2k_2+1}{k_2(k_2+1)}(L+1)\left\{\begin{array}{ccc}
        k_1 & k_2 & p_1 \\
    \frac{L}{2} & \frac{L}{2} & \frac{L}{2} \end{array}\right\}^2,
\end{eqnarray}
with the conservation law $k_1+k_2+p_1={\rm odd}~{\rm number}$ .The next correction is given explicitly by

\begin{eqnarray}\label{deltauc}
 \Gamma_2^{(3(A))} & = & \frac{1}{2}{TR}\left(-\frac{1}{2}(\Delta^{(1)})^2\right) =  -\frac{1}{4}{TR}\left[\frac{1}{\mathcal{L}^2}(\mathcal{L}\mathcal{A}+\mathcal{A}\mathcal{L})
                     \frac{1}{\mathcal{L}^2}(\mathcal{L}\mathcal{A}+\mathcal{A}\mathcal{L})\right] \nonumber \\
               & = & -\frac{1}{2}\sum_{l_1m_1}\sum_{l_2m_2}\frac{1}{l_1(l_1+1)l_2(l_2+1)}
                   \bigg[{Tr}_L[L_a,\hat{Y}_{l_1m_1}][A_a,\hat{Y}_{l_2m_2}^{+}]
           {Tr}_L[L_b,\hat{Y}_{l_2m_2}][A_b,\hat{Y}_{l_1m_1}^{+}] \nonumber\\
&+&{Tr}_L[L_a,\hat{Y}_{l_1m_1}][A_a,\hat{Y}_{l_2m_2}^{+}]
           {Tr}_L[L_b,\hat{Y}_{l_1m_1}^{+}][A_b,\hat{Y}_{l_2m_2}]\bigg].
\end{eqnarray}
This corresponds to the combination of the two diagrams displayed in Figure $6a$. Now, let us compute the two different terms separately. We have first
\begin{eqnarray}
\bar{\Gamma}_2^{(3({A}_1))}& = & -\frac{1}{2}\sum_{l_1m_1}\sum_{l_2m_2}\frac{ {Tr}_L[L_a,\hat{Y}_{l_1m_1}][A_a,\hat{Y}_{l_2m_2}^{+}]
           {Tr}_L[L_b,\hat{Y}_{l_2m_2}][A_b,\hat{Y}_{l_1m_1}^{+}]}{l_1(l_1+1)l_2(l_2+1)}
                   \nonumber\\
&=&-\frac{1}{2}(-1)^{\mu+\nu}\sum_{p_1n_1}\sum_{p_2n_2}A_{-\mu}(p_1n_1)A_{-\nu}(p_2n_2)\sum_{l_1l_2}\frac{1}{l_1(l_1+1)}\frac{1}{l_2(l_2+1)}\nonumber\\
&{\times}&\sum_{m_1m_2}(-1)^{m_1+m_2}Tr_L[L_{\mu},\hat{Y}_{l_1m_1}][\hat{Y}_{p_1n_1},\hat{Y}_{l_2-m_2}]Tr_L[L_{\nu},\hat{Y}_{l_2m_2}][\hat{Y}_{p_2n_2},\hat{Y}_{l_1-m_1}].
\end{eqnarray}
The sum over $m_1$ and $m_2$ can now be done with the help of the
identity

\begin{eqnarray}
M_1&=&(-1)^{\mu +\nu}\sum_{m_1,m_2}C_{l_1m_11\mu}^{l_1m_1+\mu}C_{l_2m_21\nu}^{l_2m_2+\nu} C_{p_1n_1l_2-m_2}^{l_1-m_1-\mu}C_{p_2n_2l_1-m_1}^{l_2-m_2-\nu}\nonumber\\
 & = &(2l_1+1)(2l_2+1)(-1)^{l_1+l_2+\nu +n_2}\sum_{km}
       C_{p_1n_11\mu}^{km}C_{p_2n_21\nu}^{k-m}\nonumber\\
&{\times}&
       \left\{\begin{array}{ccc}
        l_2 & l_1 & p_1 \\
    1 & k & l_1 \end{array}\right\} \left\{\begin{array}{ccc}
        l_1 & l_2 & p_2 \\
    1 & k & l_2 \end{array}\right\} .
\end{eqnarray}
We obtain immediately the result
\begin{eqnarray}
&&\sum_{m_1m_2}(-1)^{m_1+m_2}Tr_L[L_{\mu},\hat{Y}_{l_1m_1}][\hat{Y}_{p_1n_1},\hat{Y}_{l_2-m_2}]Tr_L[L_{\nu},\hat{Y}_{l_2m_2}][\hat{Y}_{p_2n_2},\hat{Y}_{l_1-m_1}]=\nonumber\\
&&\sqrt{l_1(l_1+1)l_2(l_2+1)}\sqrt{{\prod}_{i=1}^2(2l_i+1)(2p_i+1)}(-1)^{l_1+l_2}(L+1)[1-(-1)^{l_1+l_2+p_1}]\nonumber\\
&\times &[1-(-1)^{l_1+l_2+p_2}] \left\{\begin{array}{ccc}
        p_1 & l_1 & l_2 \\
    \frac{L}{2} & \frac{L}{2} & \frac{L}{2} \end{array}\right\} \left\{\begin{array}{ccc}
        p_2 & l_1 & l_2 \\
    \frac{L}{2} & \frac{L}{2} & \frac{L}{2} \end{array}\right\}M_1,
\end{eqnarray}
and as a consequence
\begin{eqnarray}
 \bar{\Gamma}_2^{(3({A}_1))}&=&-\frac{1}{2}\sum_{p_1n_1}\sum_{p_2n_2}A_{-\mu}(p_1n_1)A_{-\nu}(p_2n_2)(-1)^{n_1+\nu}\sum_{l_1l_2}\frac{2l_1+1}{l_1(l_1+1)}\frac{2l_2+1}{l_2(l_2+1)}(L+1)\nonumber\\
&{\times}&[1-(-1)^{l_1+l_2+p_1}][1-(-1)^{l_1+l_2+p_2}] \left\{\begin{array}{ccc}
        p_1 & l_1 & l_2 \\
    \frac{L}{2} & \frac{L}{2} & \frac{L}{2} \end{array}\right\} \left\{\begin{array}{ccc}
        p_2 & l_1 & l_2 \\
    \frac{L}{2} & \frac{L}{2} & \frac{L}{2} \end{array}\right\}f_1(lpn;\mu,\nu),\nonumber\\
\end{eqnarray}
where
\begin{eqnarray}
&&f_1=\sqrt{l_1(l_1+1)l_2(l_2+1)}\sqrt{{\prod}_{i=1}^2(2l_i+1)(2p_i+1)}\nonumber\\
&&~~~~~~~~~~~~~~\qquad\times\sum_{km}
       C_{p_1n_11\mu}^{km}C_{p_2n_21\nu}^{k-m}
       \left\{\begin{array}{ccc}
        l_2 & l_1 & p_1 \\
    1 & k & l_1 \end{array}\right\} \left\{\begin{array}{ccc}
        l_1 & l_2 & p_2 \\
    1 & k & l_2 \end{array}\right\}
\end{eqnarray}
Similarly we have
\begin{eqnarray}
\bar{\Gamma}_2^{(3({A}_2))}& = & -\frac{1}{2}\sum_{l_1m_1}\sum_{l_2m_2}\frac{{Tr}_L[L_a,\hat{Y}_{l_1m_1}][A_a,\hat{Y}_{l_2m_2}^{+}]
           {Tr}_L[L_b,\hat{Y}^{+}_{l_1m_1}][A_b,\hat{Y}_{l_2m_2}]}{l_1(l_1+1)l_2(l_2+1)}
                    \nonumber\\
&=&-\frac{1}{2}(-1)^{\mu+\nu}\sum_{p_1n_1}\sum_{p_2n_2}A_{-\mu}(p_1n_1)A_{-\nu}(p_2n_2)\sum_{l_1l_2}\frac{1}{l_1(l_1+1)}\frac{1}{l_2(l_2+1)}\nonumber\\
&{\times}&\sum_{m_1m_2}(-1)^{m_1+m_2}Tr_L[L_{\mu},\hat{Y}_{l_1m_1}][\hat{Y}_{p_1n_1},\hat{Y}_{l_2-m_2}]Tr_L[L_{\nu},\hat{Y}_{l_1-m_1}][\hat{Y}_{p_2n_2},\hat{Y}_{l_2m_2}].\nonumber\\
\end{eqnarray}
The sum over $m_1$ and $m_2$ can now be done with the help of the identity
\begin{eqnarray}
M_2&=&(-1)^{\mu +\nu}\sum_{m_1,m_2}(-1)^{m_1+m_2}C_{l_1m_11\mu}^{l_1m_1+\mu}C_{l_1-m_11\nu}^{l_1-m_1+\nu} C_{p_1n_1l_2-m_2}^{l_1-m_1-\mu}C_{p_2n_2l_2m_2}^{l_1m_1-\nu}\nonumber\\
 & = &(2l_1+1)^2(-1)^{l_1+l_2+n_1+\mu}\sum_{km}(-1)^{k}
       C_{p_1n_11\mu}^{km}C_{p_2n_21\nu}^{k-m}
       \left\{\begin{array}{ccc}
        l_2 & l_1 & p_1 \\
    1 & k & l_1 \end{array}\right\} \left\{\begin{array}{ccc}
        l_2 & l_1 & p_2 \\
    1 & k & l_1 \end{array}\right\} .\nonumber\\
\end{eqnarray}
We obtain therefore the result
\begin{eqnarray}
&&\sum_{m_1m_2}(-1)^{m_1+m_2}Tr_L[L_{\mu},\hat{Y}_{l_1m_1}][\hat{Y}_{p_1n_1},\hat{Y}_{l_2-m_2}]Tr_L[L_{\nu},\hat{Y}_{l_1-m_1}][\hat{Y}_{p_2n_2},\hat{Y}_{l_2m_2}]=\nonumber\\
&&l_1(l_1+1)(2l_2+1)\sqrt{{\prod}_{i=1}^2(2p_i+1)}(L+1)[1-(-1)^{l_1+l_2+p_1}][1-(-1)^{l_1+l_2+p_2}] \nonumber\\
&{\times}&\left\{\begin{array}{ccc}
        p_1 & l_1 & l_2 \\
    \frac{L}{2} & \frac{L}{2} & \frac{L}{2} \end{array}\right\} \left\{\begin{array}{ccc}
        p_2 & l_1 & l_2 \\
    \frac{L}{2} & \frac{L}{2} & \frac{L}{2} \end{array}\right\}M_2,
\end{eqnarray}
and hence
\begin{eqnarray}
\bar{ \Gamma}_2^{(3({A}_2))}&=&-\frac{1}{2}\sum_{p_1n_1}\sum_{p_2n_2}A_{-\mu}(p_1n_1)A_{-\nu}(p_2n_2)(-1)^{n_1+\nu}\sum_{l_1l_2}\frac{2l_1+1}{l_1(l_1+1)}\frac{2l_2+1}{l_2(l_2+1)}(L+1)\nonumber\\
&{\times}&[1-(-1)^{l_1+l_2+p_1}][1-(-1)^{l_1+l_2+p_2}] \left\{\begin{array}{ccc}
        p_1 & l_1 & l_2 \\
    \frac{L}{2} & \frac{L}{2} & \frac{L}{2} \end{array}\right\} \left\{\begin{array}{ccc}
        p_2 & l_1 & l_2 \\
    \frac{L}{2} & \frac{L}{2} & \frac{L}{2} \end{array}\right\}f_2(lpn;\mu,\nu),\nonumber\\
\end{eqnarray}
where
\begin{eqnarray}
&&f_2=l_1(l_1+1)(2l_1+1)\sqrt{{\prod}_{i=1}^2(2p_i+1)}\sum_{km}(-1)^{l_1+l_2+k}
       C_{p_1n_11\mu}^{km}C_{p_2n_21\nu}^{k-m}\nonumber\\
&&  ~~~~~~~~~~~~~~~~~~~~~~~~~~~~~~~~~~~~~~~~~~~~~~~\times  \left\{\begin{array}{ccc}
        l_2 & l_1 & p_1 \\
    1 & k & l_1 \end{array}\right\} \left\{\begin{array}{ccc}
        l_2 & l_1 & p_2 \\
    1 & k & l_1 \end{array}\right\} .\nonumber\\
\end{eqnarray}
The final answer becomes
\begin{eqnarray}
 \Gamma_2^{(3(A))}&=&-2\sum_{p_1n_1}\sum_{p_2n_2}A_{-\mu}(p_1n_1)A_{-\nu}(p_2n_2)(-1)^{n_1+\nu}\sum_{l_1l_2}\frac{2l_1+1}{l_1(l_1+1)}\frac{2l_2+1}{l_2(l_2+1)}(L+1)\nonumber\\
&{\times}&\left\{\begin{array}{ccc}
        p_1 & l_1 & l_2 \\
    \frac{L}{2} & \frac{L}{2} & \frac{L}{2} \end{array}\right\} \left\{\begin{array}{ccc}
        p_2 & l_1 & l_2 \\
    \frac{L}{2} & \frac{L}{2} & \frac{L}{2} \end{array}\right\}f^{A}(lpn;\mu,\nu),
\end{eqnarray}
where we have the conservation laws $l_1+l_2+p_1={\rm odd}~{\rm
number}$ , $l_1+l_2+p_2={\rm odd}~{\rm number}$ which means in
particular that $p_1+p_2$ can only be an even number and where
\begin{eqnarray}
&&f^{A}(lpn;\mu,\nu)=f_1+f_2=\sqrt{l_1(l_1+1)(2l_1+1)}\sqrt{{\prod}_{i=1}^2(2p_i+1)}\nonumber\\
&&\times\sum_{km}
       C_{p_1n_11\mu}^{km}C_{p_2n_21\nu}^{k-m}
\left\{\begin{array}{ccc}
        l_2 & l_1 & p_1 \\
    1 & k & l_1 \end{array}\right\}\left[
\sqrt{l_2(l_2+1)}\sqrt{2l_2+1}\left\{\begin{array}{ccc}
        l_1 & l_2 & p_2 \\
    1 & k & l_2 \end{array}\right\}\right.\nonumber\\
&&\qquad\qquad~~~~~~~~~~~~ \left. +\sqrt{l_1(l_1+1)}\sqrt{2l_1+1}(-1)^{k+l_1+l_2} \left\{\begin{array}{ccc}
        l_2 & l_1 & p_2 \\
    1 & k & l_1 \end{array}\right\}\right].
\end{eqnarray}
From this equation and from the properties of the $3j$ and $6j$
symbols it is obvious that $m=n_1+\mu=-n_2-\nu$ while $k$ takes
only the values $p_1+1$ , $p_1$ and $p_1-1$ . Hence it is also
clear that $p_2$ can only take the values $p_1$ , $p_1+2$ and
$p_1-2$ [ The values $p_1-1$ and $p_1+1$ do not contribute because
of the restriction that $p_1+p_2$ must be an even number ] . By
using the different identities on page $311$ of \cite{VKM} we can see that the function $f^{A}$ splits into two parts , a canonical gauge part plus a scalar-like part , i.e $
f^{A}=f^{A_1}+f^{A_2}$
where
\begin{eqnarray}
f^{A_1}=
\frac{1}{2}C_{p_1n_11\mu}^{p_1m}C_{p_2n_21\nu}^{p_1-m}{\eta}_{p_1}{\delta}_{p_1p_2}
\end{eqnarray}
and
\begin{eqnarray}
f^{A_2}&=&\frac{1}{2}C_{p_1n_11\mu}^{p_1-1m}C_{p_2n_21\nu}^{p_1-1-m}\big({\eta}_{p_1-1}{\delta}_{p_2,p_1}+\hat{\eta}_{p_1-1}{\delta}_{p_2,p_1-2}\big)\nonumber\\
&+&\frac{1}{2}C_{p_1n_11\mu}^{p_1+1m}C_{p_2n_21\nu}^{p_1+1-m}\big({\eta}_{p_1+1}{\delta}_{p_2,p_1}+\hat{\eta}_{p_1+1}{\delta}_{p_2,p_1+2}\big).\label{A.18}
\end{eqnarray}
The functions ${\eta}_{p_1}$  , ${\eta}_{p_1-1,p_1+1}$ and $\hat{\eta}_{p_1-1,p_1+1}$ carry all the dependence of $f^{A}$ on the internal momenta $l_1$ and $l_2$, namely
{\footnotesize{\begin{eqnarray}
&&{\eta}_{p_1}=-\frac{1}{p_1(p_1+1)}(l_2(l_2+1)-l_1(l_1+1))(l_2(l_2+1)-l_1(l_1+1)-p_1(p_1+1))\nonumber\\
&&{\eta}_{p_1+1}=\frac{1}{(p_1+1)(2p_1+3)}(s+2)(s-2p_1)(s-2l_1+1)(s-2l_2+1)\nonumber\\
&&{\eta}_{p_1-1}=\frac{1}{p_1(2p_1-1)}(s+1)(s-2p_1+1)(s-2l_1)(s-2l_2)\nonumber\\
&&\hat{\eta}_{p_1+1}=-\sqrt{\frac{(s+2)(s-2p_1)(s-2l_1+1)(s-2l_2+1)}{(p_1+1)(2p_1+3)}}\sqrt{\frac{(s+3)(s-2p_1-1)(s-2l_1+2)(s-2l_2+2)}{(2p_1+3)(p_1+2)}}\nonumber\\
&&\hat{\eta}_{p_1-1}=-\sqrt{\frac{(s+1)(s-2p_1+1)(s-2l_1)(s-2l_2)}{p_1(2p_1-1)}}\sqrt{\frac{s(s-2p_1+2)(s-2l_1-1)(s-2l_2-1)}{(2p_1-1)(p_1-1)}}.\label{Alast}
\end{eqnarray}}}
In above $s$ is defined by $s=p_1+l_1+l_2$.

In the remainder of this appendix we show explicitly that the quantum effective action $\Gamma_2^{(3A_2)}$ obtained by setting $f=f^{A_2}$ in $\Gamma_2^{(3 (A))}$ can be written in the form (\ref{5.39})  . We have
\begin{eqnarray}
 \Gamma_2^{(3A_2)}&=&-2\sum_{p_1n_1}\sum_{p_2n_2}A_{-\mu}(p_1n_1)A_{-\nu}(p_2n_2)(-1)^{n_1+\nu}\sum_{l_1l_2}\frac{2l_1+1}{l_1(l_1+1)}\frac{2l_2+1}{l_2(l_2+1)}(L+1)\nonumber\\
&{\times}&\left\{\begin{array}{ccc}
        p_1 & l_1 & l_2 \\
    \frac{L}{2} & \frac{L}{2} & \frac{L}{2} \end{array}\right\} \left\{\begin{array}{ccc}
        p_2 & l_1 & l_2 \\
    \frac{L}{2} & \frac{L}{2} & \frac{L}{2} \end{array}\right\}f^{A_2}(lpn;\mu,\nu).
\end{eqnarray}
Using (\ref{A.18}) and (\ref{Alast}) we can immediately compute
\begin{eqnarray}
\Gamma_{2}^{(3A_2)}&=&\sum_{p_1n_1}\sum_{p_2n_2}A_{-\mu}(p_1n_1)A_{-\nu}(p_2n_2)(-1)^{n_1+\nu}\nonumber\\
&{\times}&\bigg[C_{p_1n_11\mu}^{p_1-1m}C_{p_2n_21\nu}^{p_1-1-m}\bigg({\delta}_{p_1p_2}{\Lambda}^{(-)}(p_1-1)+ {\delta}_{p_2,p_1-2}{\Sigma}^{(-)}(p_1-1)\bigg)\nonumber\\
&+&C_{p_1n_11\mu}^{p_1+1m}C_{p_2n_21\nu}^{p_1+1-m}\bigg({\delta}_{p_1p_2}{\Lambda}^{(+)}(p_1+1)+{\delta}_{p_2,p_1+2}{\Sigma}^{(+)}(p_1+1)\bigg)\bigg].
\end{eqnarray}
where
\begin{eqnarray}
{\Lambda}^{(-)}(p_1-1)&=&-\sum_{l_1,l_2}\frac{2l_1+1}{l_1(l_1+1)}\frac{2l_2+1}{l_2(l_2+1)}
(L+1)\left\{\begin{array}{ccc}
        p_1 & l_1 & l_2 \\
    \frac{L}{2} & \frac{L}{2} & \frac{L}{2} \end{array}\right\}^2{\eta}_{p_1-1}\nonumber\\
{\Lambda}^{(+)}(p_1+1)&=&-\sum_{l_1,l_2}\frac{2l_1+1}{l_1(l_1+1)}\frac{2l_2+1}{l_2(l_2+1)}
(L+1)\left\{\begin{array}{ccc}
        p_1 & l_1 & l_2 \\
    \frac{L}{2} & \frac{L}{2} & \frac{L}{2} \end{array}\right\}^2{\eta}_{p_1+1},\label{540}
\end{eqnarray}
and
\begin{eqnarray}
{\Sigma}^{(-)}(p_1-1)&=&-\sum_{l_1,l_2}\frac{2l_1+1}{l_1(l_1+1)}\frac{2l_2+1}{l_2(l_2+1)}
(L+1)\left\{\begin{array}{ccc}
        p_1 & l_1 & l_2 \\
    \frac{L}{2} & \frac{L}{2} & \frac{L}{2} \end{array}\right\}\left\{\begin{array}{ccc}
        p_1-2 & l_1 & l_2 \\
    \frac{L}{2} & \frac{L}{2} & \frac{L}{2} \end{array}\right\}\hat{\eta}_{p_1-1},\label{541}\nonumber\\
\end{eqnarray}
\begin{eqnarray}
{\Sigma}^{(+)}(p_1+1)&=&-\sum_{l_1,l_2}\frac{2l_1+1}{l_1(l_1+1)}\frac{2l_2+1}{l_2(l_2+1)}
(L+1)\left\{\begin{array}{ccc}
        p_1 & l_1 & l_2 \\
    \frac{L}{2} & \frac{L}{2} & \frac{L}{2} \end{array}\right\}\left\{\begin{array}{ccc}
        p_1+2 & l_1 & l_2 \\
    \frac{L}{2} & \frac{L}{2} & \frac{L}{2} \end{array}\right\}\hat{\eta}_{p_1+1}.\label{542}\nonumber\\
\end{eqnarray}
In above we must always have $l_1+l_2+p_1$ to be an odd number . In order to rewrite the above scalar action in position space let us first introduce the following operators
\begin{eqnarray}
&&{\Delta}^{(1)}={\Delta}^{(-)}+{\Delta}^{(+)}+\bar{\Delta}^{(-)}+\bar{\Delta}^{(+)}~,~{\Delta}^{(2)}={\Delta}^{(-)}+{\Delta}^{(+)}-\bar{\Delta}^{(-)}-\bar{\Delta}^{(+)}\nonumber\\
&&{\Delta}^{(3)}={\Delta}^{(-)}-{\Delta}^{(+)}+\bar{\Delta}^{(-)}-\bar{\Delta}^{(+)}~,~{\Delta}^{(4)}={\Delta}^{(-)}-{\Delta}^{(+)}-\bar{\Delta}^{(-)}+\bar{\Delta}^{(+)},\label{543}
\end{eqnarray}
where ${\Delta}^{(-)}$ , ${\Delta}^{(+)}$ , $\bar{\Delta}^{(-)}$ and $\bar{\Delta}^{(+)}$ are defined through the equations
\begin{eqnarray}
&&{\Lambda}^{(-)}(p_1-1)=\frac{16p_1\big((L+1)^2-p_1^2\big)}{L(L+2)(2p_1-1)}{\Delta}^{(-)}(p_1-1)\nonumber\\
&&{\Lambda}^{(+)}(p_1+1)=\frac{16(p_1+1)(L+2+p_1)(L-p_1)}{L(L+2)(2p_1+3)}{\Delta}^{(+)}(p_1+1)\nonumber\\
&&{\Sigma}^{(-)}(p_1-1)= -\frac{16\sqrt{p_1(p_1-1)(L+p_1)(L+2-p_1)((L+1)^2-p_1^2)}}{L(L+2)(2p_1-1)}\bar{\Delta}^{(-)}(p_1-1)\nonumber\\
\end{eqnarray}
and
\begin{eqnarray}
{\Sigma}^{(+)}(p_1+1)&=&-\frac{16\sqrt{(p_1+1)(p_1+2)(L-p_1)(L+p_1+2)(L-p_1)(L+p_1+3)(L-p_1-1)}}{L(L+2)(2p_1+3)}\nonumber\\
&{\times}&\bar{\Delta}^{(+)}(p_1+1).
\end{eqnarray}
The above scalar action in position space is then simply given by
\begin{eqnarray}
\Delta \Gamma_{2}^{(3A_2)}&=&
Tr_L[A_a,x_a]_{+}{\Delta}^{(1)}({\cal L}^2)[A_b,x_b]_{+}+Tr_L[{\nabla}A_a,x_a]_{+}{\Delta}^{(2)}({\cal L}^2)[{\nabla}^{-1}A_b,x_b]_{+}\nonumber\\
&+&iTr_L[{\nabla}^{\frac{3}{2}}A_a,x_a]_{+}{\Delta}^{(3)}({\cal L}^2){\nabla}[{\nabla}^{\frac{3}{2}}A_b,x_b]_{+}+iTr_L[{\nabla}^{\frac{5}{2}}A_a,x_a]_{+}{\Delta}^{(4)}({\cal L}^2){\nabla}[{\nabla}^{\frac{1}{2}}A_b,x_b]_{+},\nonumber\\
\end{eqnarray}
where ${\nabla}$ is the phase operator ${\nabla}=(-1)^{\frac{\hat{N}}{2}}$ .

\bibliographystyle{unsrt}

\end{document}